\documentclass[10pt,pra,aps,floats,floatfix,twocolumn,showkeys,%
longbibliography,superscriptaddress,notitlepage,preprintnumbers,%
]%
{revtex4-1}
\usepackage[USenglish]{babel}

\usepackage{amsmath,amssymb,amsfonts}

\usepackage{graphicx}% Include figure files
\usepackage[colorlinks=true,linkcolor=blue,citecolor=red]{hyperref}

\usepackage[T1]{fontenc} %inne kodowanie (to wlaczamy ja usuniemy [MeX]{polski}
\usepackage{lmodern}

\usepackage[usenames,dvipsnames]{color}% color text
\usepackage[normalem]{ulem} % przekreslenia: \sout{Hello World}

\newcommand{\sizetwo}{0.485\textwidth}
\newcommand{\sizetwoTHREE}{0.325\textwidth}
\newcommand{\sizetwobig}{0.40\textwidth}
\newcommand{\sizethree}{0.325\textwidth}
\begin{document}
\preprint{\emph{Submitted to:} Journal of Physics: Condensed Matter (IOP Science); DOI: \href{https://doi.org/10.1088/1361-648X/aba981}{10.1088/1361-648X/aba981}}
\title{Extended Falicov-Kimball model: Hartree-Fock vs DMFT approach}
\author{Konrad Jerzy Kapcia}
\email[Corresponding author; e-mail: ]{konrad.kapcia@ifj.edu.pl}
\homepage[ORCID ID: ]{https://orcid.org/0000-0001-8842-1886}
\affiliation{Institute of Nuclear Physics, Polish Academy of Sciences, ulica W. E. Radzikowskiego 152, PL-31342 Krak\'{o}w, Poland}
\author{Romuald Lema\'nski}
\email[E-mail: ]{r.lemanski@intibs.pl}
\homepage[ORCID ID: ]{https://orcid.org/0000-0002-0852-8761}
\affiliation{Institute of Low Temperature and Structure Research, Polish Academy of Sciences, ulica Ok\'olna 2, PL-50422 Wroc\l{}aw, Poland}
\author{Marcin Jakub Zygmunt}
\email[E-mail: ]{marcin.zygmunt@us.edu.pl}
\homepage[ORCID ID: ]{https://orcid.org/0000-0001-9060-1405}
\affiliation{Institute of Mathematics, University of Silesia, ulica Bankowa 14,
PL-40007 Katowice, Poland}
\date{\today}
\begin{abstract}
In this work, we study the extended Falicov-Kimball model at half-filling within the Hartree-Fock approach (HFA) (for various crystal lattices) and compare the results obtained with the rigorous ones derived within the dynamical mean field theory (DMFT).
The model describes a system, where electrons with spin-$\downarrow$ are itinerant (with hopping amplitude $t$), whereas those with spin-$\uparrow$ are localized.
The particles interact via on-site $U$ and intersite $V$ density-density Coulomb interactions.
We show that the HFA description of the ground state properties of the model is equivalent to the exact DMFT solution and provides a qualitatively correct picture also for a range of small temperatures. 
It does capture the discontinuous transition between ordered phases at $U=2V$ for small temperatures as well as correct features of the continuous order-disorder transition.
However, the HFA predicts that the discontinuous boundary ends at the isolated-critical point (of the liquid-gas type) and it does not merge with the continuous boundary. 
This approach cannot also describe properly a change of order of the continuous transition for large $V$ as well as  various metal-insulator transitions found within the DMFT. 
\end{abstract}

% 71.27.+a	Strongly correlated electron systems,	
% 71.10.Fd	Hubbard model . electronic structure, 	
% 71.30.+h	Metal-insulator transitions and other electronic transitions
% 71.10.-w	Condensed matter . theories and models of,
% 75.20.Hr	Paramagnetism . local moment in compounds and alloys,
%\pacs{71.30.+h, 71.10.Fd, 71.27.+a, 71.10.-w}
%

\maketitle

%===================================================================================================

\section{Introduction}

Interparticle correlations in fermionic systems give rise variety of intriguing phenomena.
For example, these systems exhibit quite complex phase diagrams with, e.g., metal-insulator transitions and competition between different ordering such as spin-, charge-, orbital-order as well as superconductivity, e.g., Refs. \cite{MicnasRMP1990,ImadaRMP1998,YoshimiPRL2012,FrandsenNatCom2014,CominScience2015,%
NetoScience2015,CaiNatPhys2016,HsuNatCom2016,PelcNatCom2016,ParkPRL2017,NovelloPRL2017,%
RosciszewskiPRB2018,AvellaPRL2019,RosciszewskiPRB2019,FujiokaNatCom2019,LiuPRL2020,MallikPRL2020}.
Knowing and understanding their properties is important not only in the context of condensed matter physics, but also for physics of ultra-cold quantum gases, where intensive experimental development occurs in the recent years (for a review see, e.g., Refs. \cite{BlochRMP2008,GiorginiRMP2008,Bloch2010,GuanRMP2013,GeorgescuRMP2014,DuttaRPP2015}). 
Such systems can be used as quantum simulators of different model systems because various inter-particle interactions can be tuned very precisely.

Description of correlated electron systems requires special care and precision, because sometimes it happens that different calculation methods lead to qualitatively different results, e.g., the dependence of order-disorder transition temperature as a function of Hubbard-$U$ interaction in the attractive Hubbard model (i.e., superconducting critical temperature) \cite{MicnasRMP1990,KellerPRL2001,KogaPRA2011,ToschiNJP2005,KuleevaJETP2014} as well as in the spin-less Falicov-Kimball model (vanishing of charge order) \cite{HassanPRB2007,Lemanski2014}. 
In particular, it is believed (and in many cases it is clearly justified) that the so-called one-electron theories, as well as methods based on of a self-consistent field, are not useful for describing such systems. 
One of these methods is the Hartree-Fock approximation (HFA), which is widely used in solid state theory \cite{MicnasRMP1990,ImadaRMP1998,GeorgesRMP1996}.

The advantage of this method is its relative simplicity and the ability to describe complex systems using analytic expressions. 
However, sometimes it turns out that the accuracy of calculations is not controlled by this method. 
In particular, it cannot properly describe the Mott localization at the metal-insulator transition \cite{ImadaRMP1998,GeorgesRMP1996,AmaricciPRB2010,KapciaPRB2017}.
But there are also cases when the HFA describes correctly the behavior of interacting electron systems, particularly in the ground state, e.g., Refs. \cite{KapciaAPPA2018,LemanskiPRB2017}. 
Therefore, completely rejection of the HFA as the method for studying these systems does not seem right. 
Then, of course, the question arises: when the HFA correctly describes a given system?

To find the answer to this question, in this work we analyze the extended Falicov-Kimball model (EFKM) \cite{DongenPRL1990,DongenPRB1992,LemanskiPRB2017,KapciaPRB2019,KapciaPRB2020}  in a wide range of interaction parameters and temperature by using the HFA and compare the results with those obtained within the dynamical mean field theory (DMFT). 
We consider the case of the Bethe lattice in the limit of large dimensions, when the DMFT is the exact method \cite{MullerHartmannZPB1989,LemanskiPRB2017,KapciaPRB2019,KapciaPRB2020}.
Thanks to this, we can fix ranges of the model parameters for which the HFA gives results close to exact, as well as those for which exact results are qualitatively different from those obtained using the HFA. 
In other words, we determine the ranges of HFA applicability for the tested model  in a controlled way on a basis of exact DMFT calculations.

In general, the HFA fails in finite temperature when on-site Coulomb interaction $U$ is present. 
However, at $T = 0$, analytic expressions for electron density and energy obtained using HFA and DMFT are proven to be equivalent (see Appendix \ref{sec:appHFAvsDMFT}). 
Therefore, it is natural to expect that also in a low temperature range and/or for small values of the interaction parameter $U$  we will get similar results when we use the HFA and the DMFT. 
And indeed, our calculations confirm this hypothesis.

The present work is organized in the following way.
Section \ref{sec:modelandmethod} describes the model investigated (Sec. \ref{sec:model}) and the method used (Sec. \ref{sec:HFA}, includes also equation at $T=0$ and $T>0$ for the order parameters, the free energy as well as for the transition temperature).
Section \ref{sec:numresults} is devoted to presentation of numerical results such as a phase diagram of the model and dependencies of various thermodynamical quantities (Sec. \ref{sec:phasediagHFA}) and a comparison of these findings with the rigorous results (Sec. \ref{sec:validityHFA}).
In Section \ref{sec:conclusions}, the conclusions and final remarks are presented.
The appendixes are devoted to a rigorous proof of the fact that the HFA is an exact theory for the model at $T=0$ (Appendix \ref{sec:appHFAvsDMFT})  and to an analysis of the equations obtained for a very particular case of $U=2V$ (Appendix \ref{sec:appU2Vtransition}).

\section{Model and method}
\label{sec:modelandmethod}

\subsection{Extended Falicov-Kimball model at half-filling}
\label{sec:model}

The Hamiltonian of the EFKM (cf. also Refs. \cite{DongenPRL1990,DongenPRB1992,LemanskiPRB2017,KapciaPRB2019,KapciaPRB2020}) has the following form  
\begin{eqnarray}
\label{eq:ham}
H & = & 
\frac{t}{\sqrt{z}}\sum_{\left\langle i,j\right\rangle}{ \left( \hat{c}^{\dag}_{i,\downarrow}\hat{c}_{j,\downarrow} + \hat{c}^{\dag}_{j,\downarrow}\hat{c}_{i,\downarrow} \right) }
 + U\sum_i \hat{n}_{i,\uparrow} \hat{n}_{i,\downarrow} \\
 & + & \frac{2V}{z} \sum_{\left\langle i,j\right\rangle,\sigma,\sigma'} \hat{n}_{i,\sigma}\hat{n}_{j,\sigma'} - \sum_{i,\sigma} \mu_{\sigma}\hat{n}_{i,\sigma}, \nonumber
\end{eqnarray}
where $\hat{c}^{\dag}_{i,\downarrow}$ ($\hat{c}_{i,\downarrow}$) denotes creation(annihilation) of fermion (electron) with spin $\sigma$ ($\sigma \in \{ \downarrow, \uparrow \}$) at site $i$ and $\hat{n}_{i,\sigma} = \hat{c}^{\dag}_{i,\downarrow} \hat{c}_{i,\downarrow}$. 
$U$ and $V$ denote on-site and intersite nearest-neighbor, respectively,  density-density Coulomb interactions.
$\sum_{\langle i,j \rangle }$ indicates summation over the nearest-neighbor pairs.
$\mu_{\sigma}$ is the site-independent chemical potential for electrons with spin $\sigma$.
In this work, we consider the case of half-filling, i.e., $\mu_{\sigma} = (U + 4V)/2$ for both directions of the spin.
The denotation used are the same as those used in Refs. \cite{DongenPRB1992,LemanskiPRB2017,KapciaPRB2019,KapciaPRB2020}.

A review of properties of the standard Falicov-Kimball model (FKM) [called also as the spin-less Falicov-Kimball model, i.e., $V=0$ case of model (\ref{eq:ham}), particularly in the infinite dimension limit] can be found, e.g., in Refs. \cite{FalicovPRL1969,KennedyPhysA1986,LiebPhysA1986,BrandtMielsch1989,BrandtMielsch1990,BrandtMielsch1991,%
BrandtUrbanek1992,FreericksPRB1999,Jedrzejewski2001,ChenPRB2003,FreericksRMP2003,%
FreericksBook2006,HassanPRB2007,Lemanski2014,Lemanski2016,%
Krawczyk2018,ZondaPRB2019,AstleithnerPRB2020}.
One should note that other extensions of the Falicov-Kimball model such as an explicit local hybridization, a level splitting, various nonlocal Coulomb interaction, correlated and extended hoppings, or a consideration of a larger number of localized states are also possible, e.g., Refs. \cite{BrydonPRB2005,YadavEPL2011,Farkasovsky2015,HamadaJPhysSocJap2017,Farkasovsky2019}
and extensive lists of references, which can be found in the reviews (cf. Refs. \cite{FreericksRMP2003,Gruber1996,Jedrzejewski2001}).

From the historical perspective, the standard FKM appears in Hubbard's original work \cite{HubbardPRSL1963} and it was analyzed as an approximation of the full Hubbard model in Ref. \cite{HubbardPRSLA1964}.
Next, it was proposed for a description of transition metal oxides \cite{FalicovPRL1969} as well as a model for crystallization \cite{KennedyPhysA1986,LiebPhysA1986}.
Moreover, the FKM can describe some anomalous properties of rare-earth compounds with an isostructure valence-charge transition, such as Yb$_{1-x}$Y$_{x}$InCu$_{4}$ or EuNi$_{2}$(Si$_{1-x}$Ge$_{x}$)$_{2}$ materials \cite{ZlaticAPPB2001,ZlaticAPPB2003}.
It can also successfully  describe electron Raman scattering features in, e.g., SmB$_{6}$ and FeSi  in the insulting phase \cite{FreericksDevereauxPRB2001,FreericksDevereauxCMP2001,FreericksDevereauxAPPB2001}.
The FKM can be also applied to the pressure-induced isostructural metal-insulator transition in NiI$_{2}$ \cite{PasternakPRL1990,FreericksFalicovPRB1992,ChenPRL1993}.
Other  example, where the FKM can give prediction on real systems, is the field of Josephson junctions (e.g., in Ta$_{x}$N) \cite{FreericksNikolicPRB2001,FreericksNikolicIJMPB2002,FreericksNikolicPRB2003}.
The model was used also for an explanation of behavior of colossal magneto-resistance materials \cite{AllubAlascioPRB1997,LetfulovPRB1999}.
However, one should underline that, in reality, not only onsite interactions occurs, thus the inclusion of intersite repulsion, as it is done in the EFKM [Eq. (\ref{eq:ham})], could give a better insight into the physics of real materials.

In this paper, we use  mainly the semi-elliptic density of states, which is the DOS of non-interacting particles on the Bethe lattice with the coordination number $z\rightarrow \infty$ \cite{GeorgesRMP1996,FreericksRMP2003}. 
In addition, we use the gaussian DOS, which is specific one for the model of tight-binding electrons on the $d$-dimensional hypercubic lattice \cite{MetznerVollhardtPRL1989,MullerHartmannZPB1989,GeorgesRMP1996}, and the lorentzian DOS, which can be realized with a hopping matrix involving the long-range hopping \cite{Georges1992}.
For a comparison we also use the rectangular DOS.
The explicit forms for the used DOSs are as follows:
(i) the semi-elliptic DOS: $D_{S-E}(\varepsilon)=(2\pi t^2)^{-1}\sqrt{4t^2-\varepsilon^2}$ for $|\varepsilon| \leq 2t $ and $D_{S-E}(\varepsilon) = 0$ for $|\varepsilon | > 2t $; 
(ii) the gaussian DOS: $D_{G}(\varepsilon) = \left( t \sqrt{2 \pi } \right)^{-1} \exp \left[ - \varepsilon^2/\left(2t^2\right) \right]$;
(iii) the lorentzian DOS: $D_{L}(\varepsilon) = t \left[\pi \left( \varepsilon^2 + t^2 \right) \right]^{-1}$;
(iv) the rectangular DOS: $D_{R}(\varepsilon)=1/(4 t)$ for $|\varepsilon |\leq 2t $ and $D_{R}(\varepsilon) = 0$ for $|\varepsilon | > 2t $.
In all these cases, the half-bandwidth is defined as $2t$. 
In the rest of the paper, we take $t$ as an energy unit.

\subsection{Hartree-Fock approach}
\label{sec:HFA}

Let us consider the model on a bipartite (alternate) lattice, i.e., on the lattice which can be divided into two sublattices (denoted by $\alpha=A,B$) in such a way that all nearest-neighbors of a site from one sublattice belong to the other sublattice.

The Hamiltonian (\ref{eq:ham}) is treated within the standard broken symmetry mean-field Hartree-Fock approach \cite{Penn1966,MicnasRMP1990,ImadaRMP1998,RobaszkiewiczPRB1999} using the Bogoliubov transformation \cite{Bogoljubov1958,Valatin1958} and restricting only to Hartree terms \cite{MullerHartmannZPB1989}.
Namely, we use the following decoupling of two-particle operators:
\begin{equation*}
\hat{n}_{i\sigma}\hat{n}_{j\sigma'}= \hat{n}_{i\sigma} \left\langle \hat{n}_{j\sigma'} \right\rangle + \left\langle \hat{n}_{i\sigma} \right\rangle \hat{n}_{j\sigma'} - \left\langle \hat{n}_{i\sigma}\right\rangle \left\langle \hat{n}_{j\sigma'} \right\rangle,
\end{equation*}  
where $\langle \hat{A} \rangle $ denotes the average value of the operator $\hat{A}$ (in the thermodynamic meaning).
Note that this decoupling is an exact one for intersite term (i.e., $i \neq j$) in the limit of large dimension \cite{MullerHartmannZPB1989}.
The interaction part of the Hamiltonian (\ref{eq:ham}) including $U$-, $V$- and $\mu$- terms at the half-filling (i.e., $\sum_{i}\langle \hat{n}_{i,\sigma}\rangle = N/2$, $N$ -- the number of the lattice sites) can be written in the form:
\begin{equation*}
\hat{H}^{int}_{MF} = \sum_{i,\sigma} W_{\sigma} \hat{n}_{i,\sigma} \exp{(\mathbf{i} \vec{Q}\cdot \vec{R}_{i})} + NC,
\end{equation*}
where $C=\frac{1}{4}\left\{ -U \left( 1 + \Delta_{\uparrow} \Delta_{\downarrow} \right) + V \left[ \left ( \Delta_{\uparrow} + \Delta_{\downarrow} \right)^{2} - 4\right] \right\}$, $\exp{\left(\mathbf{i} \vec{Q}\cdot \vec{R}_{i}\right)} = \pm 1$ if $i\in A, B$, respectively.
$\vec{Q}$ is a half of the largest reciprocal lattice vector in the first Brillouin zone, $\vec{R}_i$ indicates the location of $i$-th site, and $W_\sigma=U\Delta_{\sigma}/2 - V(\Delta_{\uparrow}+\Delta_{\downarrow})$.
Parameters $\Delta_{\downarrow}$ and $\Delta_{\uparrow}$ are defined as differences between average occupation of sublattices by itinerant and localized electrons, respectively (cf. Refs. \cite{DongenPRL1990,DongenPRB1992,LemanskiPRB2017,KapciaPRB2019,KapciaPRB2020}). Namely,
\begin{equation}
\label{eq:orderparameters}
\Delta_{\sigma} = \tfrac{2}{N} \sum_i \left \langle \hat{n}_{i,\sigma} \right\rangle \exp{\left(\mathbf{i} \vec{Q}\cdot \vec{R}_{i}\right)} =  n^{A}_{\sigma} - n^{B}_{\sigma},
\end{equation}
where $n^{\alpha}_{\sigma} = \langle \hat{n}_{i,\sigma} \rangle$ for any $i \in \alpha$, where
$\alpha=A,B$ denotes the sublattice index.
Now, the Hamiltonian (\ref{eq:ham}) in the HFA can be written in terms of 
two sublattice operators in the reduced Brillouin zone (RBZ)
as $\hat{H}_{MF}=\sum_{\vec{k},\sigma}\hat{\Phi}_{\vec{k},\sigma}^\dagger \mathbb{H}_{\vec{k},\sigma} \hat{\Phi}_{\vec{k},\sigma}+NC$, where 
$\hat{\Phi}_{\vec{k},\sigma}^\dagger = \left( \hat{c}^{\dagger}_{A,\vec{k},\sigma}, \hat{c}^{\dagger}_{B,\vec{k},\sigma} \right)$ are the Nambu spinors,
$\hat{c}^{\dagger}_{\alpha,\vec{k},\sigma}$ and $\hat{c}_{\alpha,\vec{k},\sigma}$ are fermion operators defined by the discrete Fourier transformation:
\begin{eqnarray*}
\hat{c}^{\dagger}_{i,\sigma} & = & \sqrt{\frac{2}{N}}\sum_{\vec{k}} \hat{c}^{\dagger}_{\alpha,\vec{k},\sigma} \exp \left(\mathbf{i}\vec{k} \cdot \vec{R}_i\right) \quad  \textrm{for} \  i\in \alpha;\\
\hat{c}^{\dagger}_{\alpha,\vec{k},\sigma} & = & \sqrt{\frac{2}{N}} \sum_{i\in \alpha} \hat{c}^{\dagger}_{i,\sigma} \exp \left(-\mathbf{i}\vec{k} \cdot \vec{R}_i\right). 
\end{eqnarray*}
Matrix   $\mathbb{H}_{\vec{k},\sigma}$ has the form
\begin{equation*}
\mathbb{H}_{\vec{k},\sigma} = \left(
\begin{array}{cc}
W_{\sigma} & \epsilon_{\vec{k},\sigma}\\
\epsilon_{\vec{k},\sigma} & -W_{\sigma}
\end{array}\right)
\end{equation*} 
with $\epsilon_{\vec{k},\uparrow} = 0$ and $\epsilon_{\vec{k},\downarrow} = (t/\sqrt{z}) \sum_{m} \exp{(-\mathbf{ i} \vec{k} \cdot \vec{\delta}_{m})} $, $\vec{\delta}_{m}$ defines the locations of the nearest-neighbor sites in the unit cell consisting of two lattice sites and the sum is done over all nearest neighbors.
All summations over momentum $\vec{k}$ are performed over the RBZ (there is $N/2$ states in the RBZ for each sublattice per spin).
The eigenvalues $E_{\vec{k},\sigma}^{\pm}$ of $\mathbb{H}_{\vec{k},\sigma}$ are given by $E_{\vec{k},\sigma}^{\pm} = \pm \sqrt{\epsilon_{\vec{k},\sigma}^{2} + W_{\sigma}^2}$.
Thus, the full mean-field Hamiltonian can be expressed in the quasi-particle excitation as 
$\hat{H}_{MF}^{'} = \sum_{\vec{k},\sigma,r=\pm} E_{\vec{k},\sigma}^{\pm} \gamma^{\dagger}_{\vec{k},\sigma,r} \gamma _{\vec{k},\sigma,r} + NC$, where $\gamma^{\dagger}_{\vec{k},\sigma,r}$ are creation operators of the fermionic quasiparticles. 
The grand canonical potential $\Omega$ (per site) of the system is defined as 
$\Omega = -1/(N\beta) \ln \{ \textrm{Tr}[\exp(-\beta \hat{H}_{MF})] \}$. 
One gets $\Omega= - 1/(2N\beta) \sum_{\vec{k},\sigma,r=\pm} \ln \left[1+ \exp \left(-\beta E_{\vec{k},\sigma}^{r}\right) \right] + C$. 
The free energy $F = \Omega +  \sum_{i,\sigma} \mu_{\sigma} \langle \hat{n}_{i,\sigma} \rangle$ in the half-filled case is derived as $F= - 1/(N\beta) \sum_{\vec{k},\sigma} \ln \left[2 \cosh\left(\beta E_{\vec{k},\sigma}^{+}/2\right) \right] + C + (U+4V)/2$. 
Finally, one obtains the free energy of the EFKM (per site) in the following form   
\begin{eqnarray}
\label{eq:freeener}
F & = & \frac{1}{4}\left\{ U \left( 1 - \Delta_{\uparrow} \Delta_{\downarrow} \right) + V \left[ 4 + \left ( \Delta_{\uparrow} + \Delta_{\downarrow} \right)^{2} \right] \right\} \\
& - &  \frac{1}{\beta} \ln \left[ 2 \cosh \left( \beta B / 4 \right) \right] \nonumber \\
& - &  \frac{1}{\beta} \int_{-\infty}^{+\infty}{ 
D(\varepsilon) \ln 
\left[ 2 \cosh \left( \tfrac{\beta}{2} \sqrt{\varepsilon^2 + A^{2}} \right) \right] d \varepsilon 
}, \nonumber
\end{eqnarray}
where $A = - U \Delta_{\uparrow}/2 + V \left( \Delta_{\uparrow} + \Delta_{\downarrow}\right) $,
$B = - U \Delta_{\downarrow} + 2 V \left( \Delta_{\uparrow} + \Delta_{\downarrow} \right)$,
%$\Delta_{\sigma} = n_{A,\sigma} - n_{B,\sigma}$, $n_{\alpha,\sigma} = (2/L) \sum_{i\in \alpha} \left\langle \hat{n}_{i,\sigma} \right\rangle $, 
and $\beta=1/(k_{B}T)$. 
$T$ denotes temperature and $k_B$ is the Boltzmann constant.
$D(\varepsilon)$ is the non-interacting density of states (DOS), which an explicit form is dependent on the particular lattice on which model (\ref{eq:ham}) is considered (as described in Sec. \ref{sec:model}).

After some straightforward transformations, one also gets the self-consistent equations for parameters $\Delta_{\downarrow}$ and $\Delta_{\uparrow}$ in the following form:
\begin{eqnarray}
\Delta_{\downarrow} & = & A \int_{-\infty}^{+\infty}{ 
D(\varepsilon) 
\frac{\tanh \left( \tfrac{\beta}{2} \sqrt{\varepsilon^{2} + A^{2}} \right)}{\sqrt{\varepsilon^2 + A^{2}}} 
d\varepsilon,
}
\label{eq:DeltaDown}\\
\Delta_{\uparrow} & = & \tanh \left(\beta  B  / 4 \right).  
\label{eq:DeltaUp}
\end{eqnarray}
Value $\Delta_{F}(\varepsilon) = 2 A$ can be also interpreted as an energy gap at the chemical potential (the Fermi level at $T=0$) for Bogoliubov quasiparticles \cite{LemanskiPRB2017}. 

Let us also notice, that Eqs. (\ref{eq:DeltaDown})--(\ref{eq:DeltaUp}) could be also obtained from the conditions $\partial F / \partial \Delta_{\sigma} = 0 $ (formally, with an exclusion of  $U=0$ and $U=4V$).
These are, however, only the necessary conditions for the extremum value of (\ref{eq:freeener}) with respect to $\Delta_{\downarrow}$ and $\Delta_{\uparrow}$. 
Thus, the solutions of (\ref{eq:DeltaDown})--(\ref{eq:DeltaUp}) can correspond to a local minimum, a local maximum, or a point of inflection of $F(\Delta_{\downarrow},\Delta_{\uparrow})$. 
In addition, the number of minima can be larger than one (particularly for small $T>0$), so it is very important to find the solution which corresponds to the global minimum of (\ref{eq:freeener}).

Note that if pair $(\Delta_\downarrow, \Delta_\uparrow)$ is a solution of set (\ref{eq:DeltaDown})--(\ref{eq:DeltaUp}) then pair $( -\Delta_\downarrow, -\Delta_\uparrow)$ is also a solution of the set [Eqs. (\ref{eq:DeltaDown})--(\ref{eq:DeltaUp}) do not change their forms if one does substitution $\Delta_{\downarrow} \rightarrow -\Delta_{\downarrow}$ and $\Delta_{\uparrow} \rightarrow -\Delta_{\uparrow}$]. 
Also from (\ref{eq:freeener}), one gets that $F(U,V,\Delta_{\downarrow},\Delta_{\uparrow})=F(U,V,-\Delta_{\downarrow},-\Delta_{\uparrow})$ at any $\beta$. 
Thus, we can further restrict ourselves to find the solutions of (\ref{eq:DeltaDown})--(\ref{eq:DeltaUp}) only with $\Delta_{\uparrow} \geq 0$ (parameter $\Delta_{\downarrow}$ can be of any sign). 
These properties are connected with an equivalence of both sublattices of an alternate lattice.

These two parameters can be connected with charge polarization $n_Q$ and staggered magnetization $m_Q$ by relation: $n_Q = \left( \Delta_{\uparrow} + \Delta_{\downarrow} \right)$ and $m_Q = \left( \Delta_{\uparrow} - \Delta_{\downarrow} \right)$, which create a different, but totally equivalent, set of parameters ($n_Q \geq 0$ and $m_Q \geq 0$, because of assumed $\Delta_{\uparrow} \geq 0$ and relation $\Delta_{\uparrow} \geq \Delta_{\downarrow} $ founded) \cite{LemanskiPRB2017,KapciaPRB2019,KapciaPRB2020}.    
These quantities define various phases occurring in the system.
A solution with $\Delta_{\downarrow}=\Delta_{\uparrow}=0$ ($n_Q=m_Q=0$) corresponds to the nonordered (NO) phase.
In the ordered phases, $\Delta_{\uparrow}\neq 0 $ or $\Delta_{\downarrow}\neq 0 $ ($n_Q \neq 0 $ or $ m_Q \neq 0 $).
We distinguish two such phases ($\Delta_{\uparrow}>0$ assumed): 
(i) the CO phase, where charge order dominates, i.e., $\Delta_{\downarrow}>0$ ($n_Q>m_Q$) and
(ii) the AF phase, where antiferromagnetic order is dominant, i.e., $\Delta_{\downarrow}<0$ ($n_Q<m_Q$). 
A very special case of $\Delta_{\uparrow}>0$ and $\Delta_{\downarrow}=0$ (i.e., $n_Q = m_Q > 0$) occurring for $U=2V$ is discussed in detail in Sec. \ref{sec:phasediagHFA}. 
Note that, in both CO and AF phases, the long-range order breaks the same translation symmetry.

\begin{figure*}[t]
	\includegraphics[width=\sizetwo]{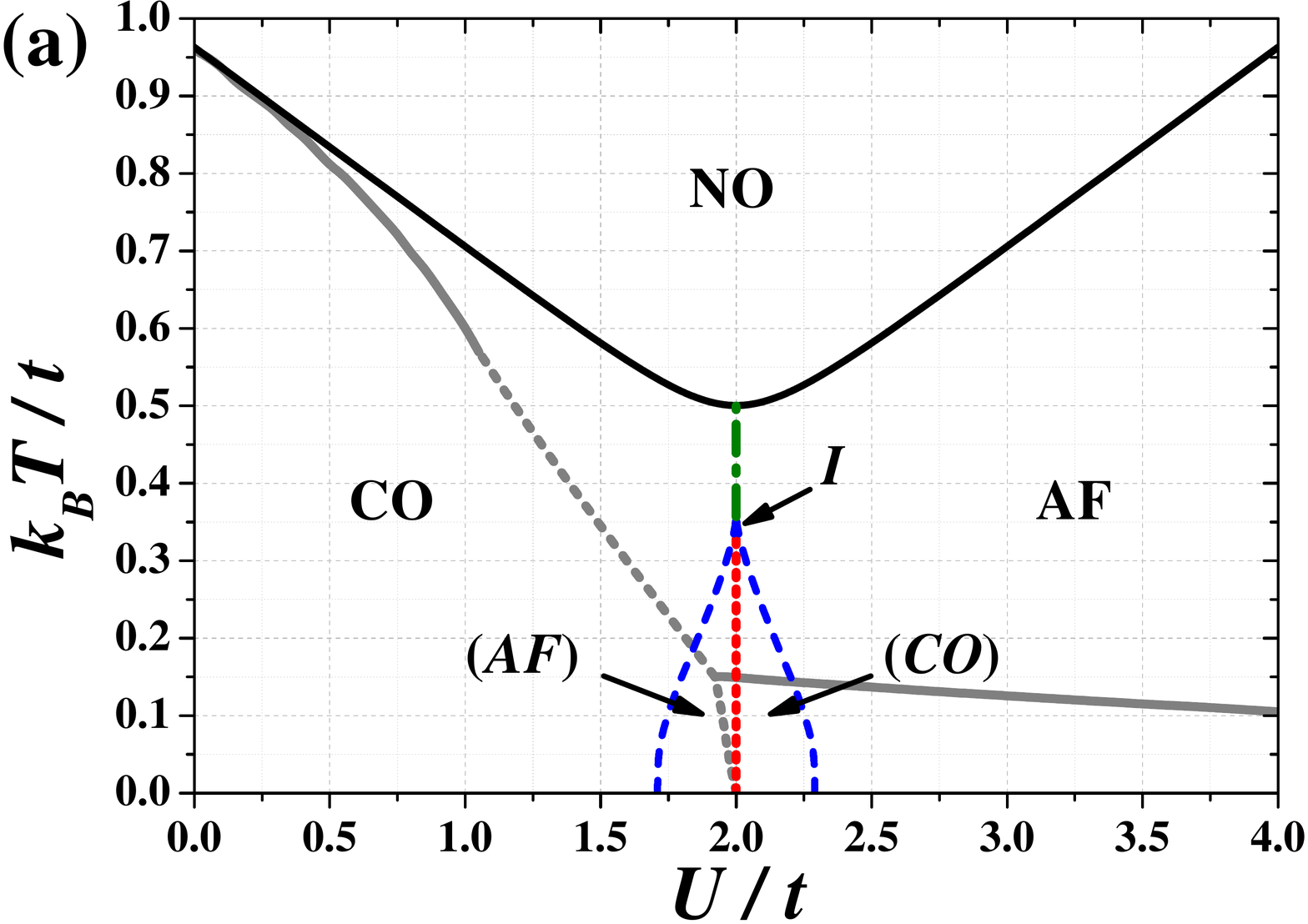}
	\includegraphics[width=\sizetwo]{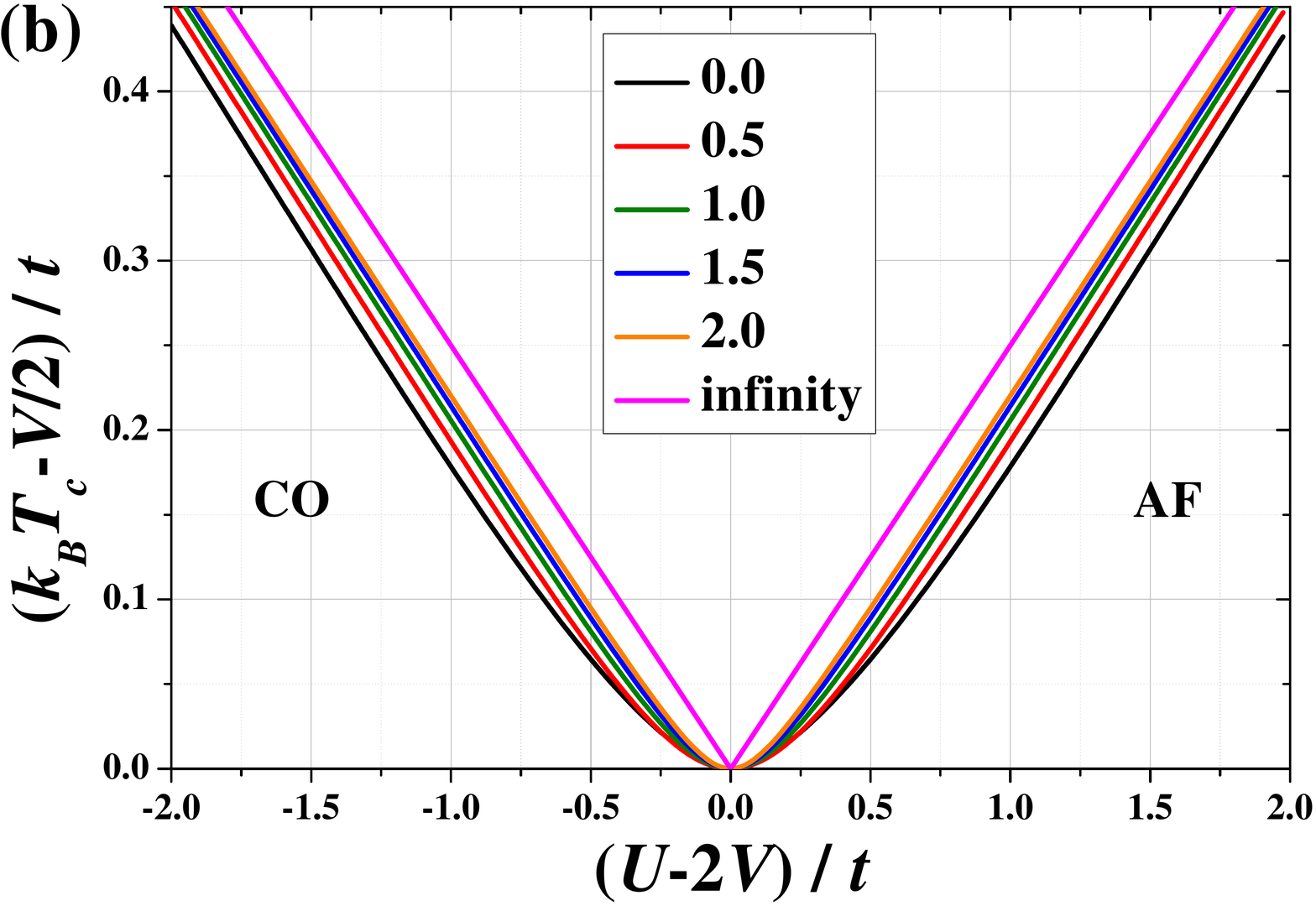}
	\caption{(a) Exemplary $k_BT/t$--$U/t$ phase diagram for $V/t=1.0$.
		Solid black and dotted red lines denote second and first order transitions.
		CO, AF, and NO denotes phases with dominant charge-order, 
		antiferromagnetic order, and nonordered phases, respectively.
		Dashed blue lines denote the regions of occurrence of (ordered) metastable phases 
		(names in the brackets) in the neighborhood of discontinuous transition.
		$I$ indicates the isolated-critical point located at $T_{c}^{*}$.
		The dash-dotted green line denotes the points, 
		where $\Delta_{\downarrow}$ changes its sign continuously (it is not a transition, rather a crossover).
		(b) The continuous order-disorder transition temperature $T_c$ 
		as a function of model parameters for several values of  intersite interaction $V/t$ 
		(as labeled, increasing from the bottom).
		Line $V/2 + |U - 2V|/4$ is an asymptotic expression for $k_{B}T_{c}$ for $V\rightarrow+\infty$ 
		or $U\rightarrow \pm \infty$ (equivalent with $t \rightarrow 0$ limit).
		All results obtained within the HFA for the semi-elliptic DOS.
		Additionally, on panel (a), solid and dashed grey lines correspond to continuous 
		and discontinuous transitions found in the DMFT, taken from Ref. \cite{KapciaPRB2019}. 
		}
	\label{fig:HFAexphasediagram}
\end{figure*}

Please also note that the HFA for the intersite term restricted only to the Hartree terms is an exact approach in the limit of large dimensions \cite{MullerHartmannZPB1989}, but for the onsite term in a general case the DMFT needs to be used \cite{GeorgesRMP1996,ImadaRMP1998}.
However, the HFA can work correctly in the ground state for some particular models and phases (cf. Appendix \ref{sec:appHFAvsDMFT} and Ref. \cite{LemanskiPRB2017}).

\subsubsection{Expressions for the ground state}
\label{sec:expressionsGS}

From above equations, for $\beta \rightarrow + \infty$, one gets $\Delta_{\uparrow}=1$ and derives the expressions in the ground state (i.e., at $T=0$) for parameter $\Delta_{\downarrow}$ in the form of
\begin{equation}
\label{eq:DeltadownGS}
\Delta_{\downarrow}= A_0 \int_{-\infty}^{+\infty}{ \frac{D(\varepsilon)d \varepsilon}{\sqrt{4 \varepsilon^{2} + A_{0}^{2}}} }
\end{equation}
as well as for the free energy per site [which is equal to the internal energy of the systems (per site) in the ground state] as
\begin{eqnarray}
\label{eq:freeenergyGS}
F_{0} 
& = &  E_{0} -  \frac{1}{4} \int_{-\infty}^{+\infty}{ D(\varepsilon) \sqrt{4 \varepsilon^{2} + A_{0}^{2}} d \varepsilon },
\end{eqnarray}
where
$A_0$ and $E_0$ are expressed by 
$A_0  = 2V \left(1+\Delta_{\downarrow}\right) - U$ and  $E_{0} = \tfrac{1}{4}\left[ U + V \left( 3 + \Delta_{\downarrow}^{2} \right) \right]$.
Note that these expressions are exactly the same as a rigorous solution in the limit of large dimensions for the FKM (obtained within the DMFT, at least for the Bethe lattice;  cf. Appendix \ref{sec:appHFAvsDMFT}).

\begin{figure*}
	\includegraphics[width=\sizetwo]{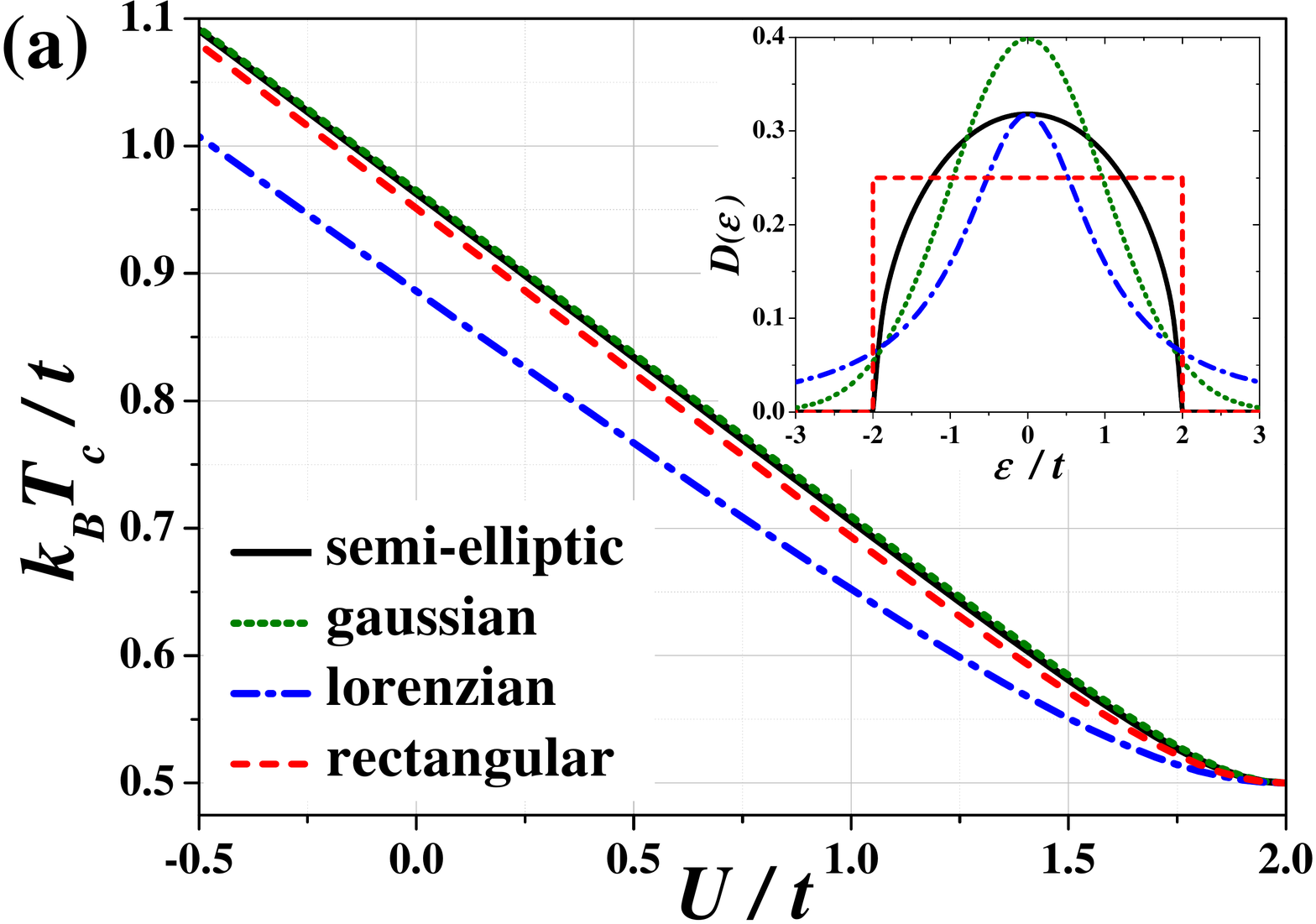}
	\includegraphics[width=\sizetwo]{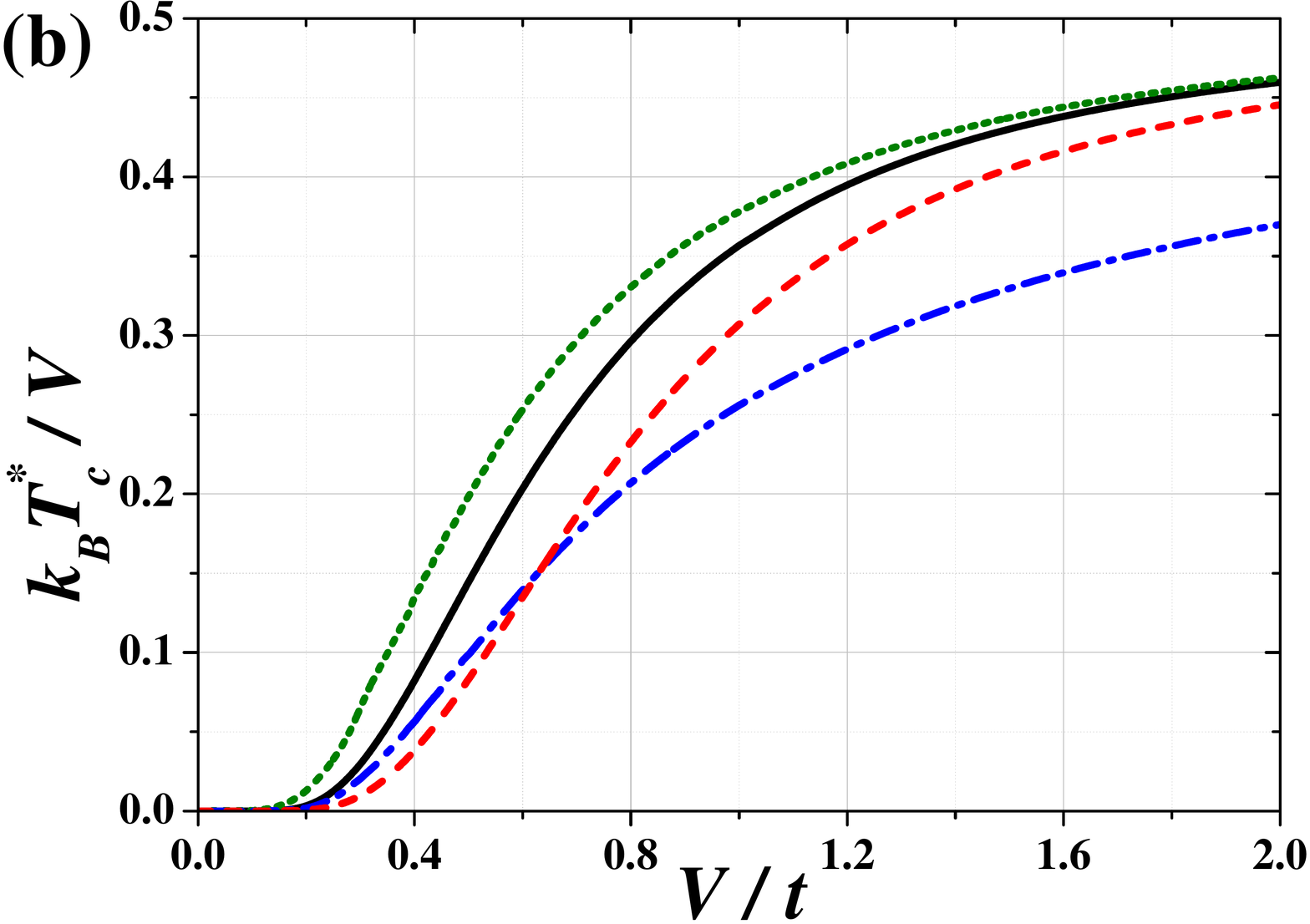}
	\caption{
		(a) The dependence of order-disorder transition temperature $T_{c}$ 
		as a function of $U/t$ for $V/t=1.0$ and different DOSs.
		Inset: shapes of the DOSs used in calculations.
		(b) The dependence of location $T_{c}^{*}$ of isolated critical point $I$ 
		as a function of $V/t$ for and different DOSs.
		The solid, dotted, dot-dashed, and dashed lines correspond to 
		the semi-elliptic, gaussian, lorentzian, and rectangular DOSs, respectively 
		(all results obtained within the HFA). 
	}
	\label{fig:HFATcdifferentDOS}
\end{figure*}

\subsubsection{Equation for temperature of the continuous order-disorder transition}
\label{sec:equationTc}

Assuming the continuous vanishing of both parameters $\Delta_{\sigma} \rightarrow 0$ at  $T_c$, one can obtain from Eqs. (\ref{eq:DeltaDown})--(\ref{eq:DeltaUp}) the following equations determining the temperature $T_c$ of the continuous order-disorder transition   
\begin{equation}
\label{eq:crittemp}
\frac{4}{\beta_c}=\left(-U + 2V \right)^2 \left( \frac{2}{I_{c}\left(\beta_{c}\right)}- 2 V \right)^{-1} + 2V,
\end{equation}
where integral $I_{c}\left(\beta_{c}\right)$ is defined as
\begin{equation}
\label{eq:crittempIntegral}
I_{c}\left(\beta_{c}\right) = \int_{-\infty}^{+\infty}{ \frac{D(\varepsilon)}{\varepsilon} \tanh{\left( \frac{\beta_c \varepsilon}{2} \right)} d\varepsilon}
\end{equation}
and $\beta_{c}^{-1} = k_{B}T_{c}$.
To obtain above relation we used that $\lim_{x \rightarrow 0} \left[ \tanh (k x) /x \right]= k$ (where $k\in \mathbb{R}$) and assuming that $\Delta_{\uparrow} \propto \ \Delta_{\downarrow} $ near $T_{c}$.
It turns out that Eq. (\ref{eq:crittemp}), in the range $0<U<4V$, has two solutions, but only the solution $ k_{B}T_{c} \geq V/2$ is physical and coincides with the order-disorder phase boundaries determined by comparison of free energies of different solutions and presented in Sec. \ref{sec:numresults}.
It turns out that the other one (that smaller than $V/2$) corresponds to vanishing of a local maximum of $F(\Delta_{\downarrow},\Delta_{\uparrow})$ at $(0,0)$ (an unstable solution). 
Note also that, for $U=2V$, assumption $\Delta_{\uparrow} \propto \ \Delta_{\downarrow} $ cannot be fulfilled.
This particular case is studied in detail at Appendix \ref{sec:appU2Vtransition}.

In addition, for $t\rightarrow 0$, $D(\varepsilon)\rightarrow \delta(\varepsilon)$ (for any of explicit forms used further in the paper), $I_c(\beta_{c}) \rightarrow \beta_{c}/2$ and $2k_{B}T_{c} = V + |V - U/2|$.
Here, $\delta(\varepsilon)$ denotes the Dirac function (distribution).
It is clearly seen that this result is different than the rigorous result obtained at atomic limit of the EFKM (in infinite dimension limit) \cite{MicnasPRB1984,KapciaPhysA2016}.
In this limit, the HFA coincides with the exact result only for  $U=0$ and one gets that $k_{B}T_{c}=V$.

\section{Numerical results}
\label{sec:numresults}

\subsection{Phase diagram of the model within the Hartree-Fock approximation}
\label{sec:phasediagHFA}

\begin{figure*}	
    \includegraphics[width=\sizetwoTHREE]{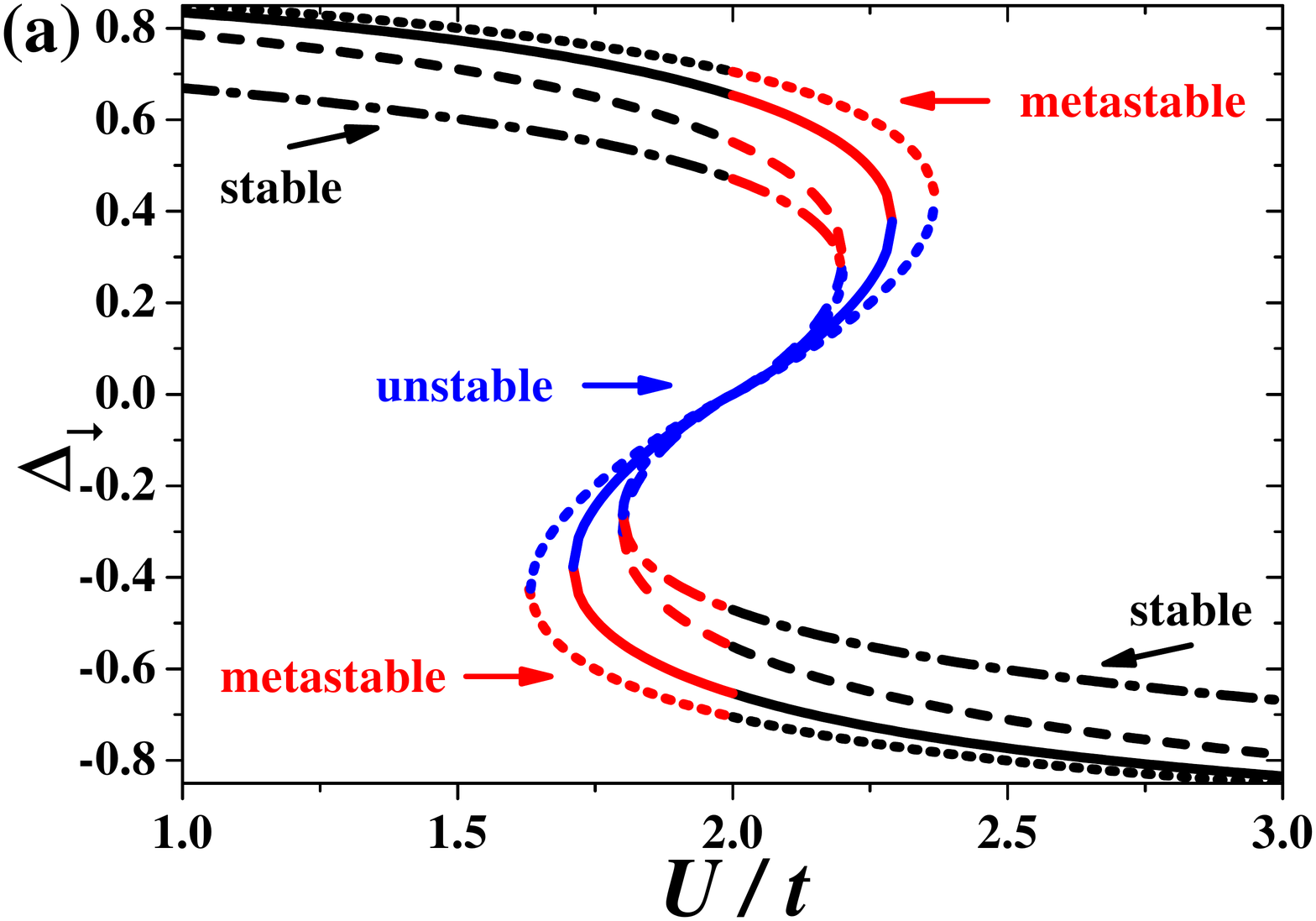}
   \includegraphics[width=\sizetwoTHREE]{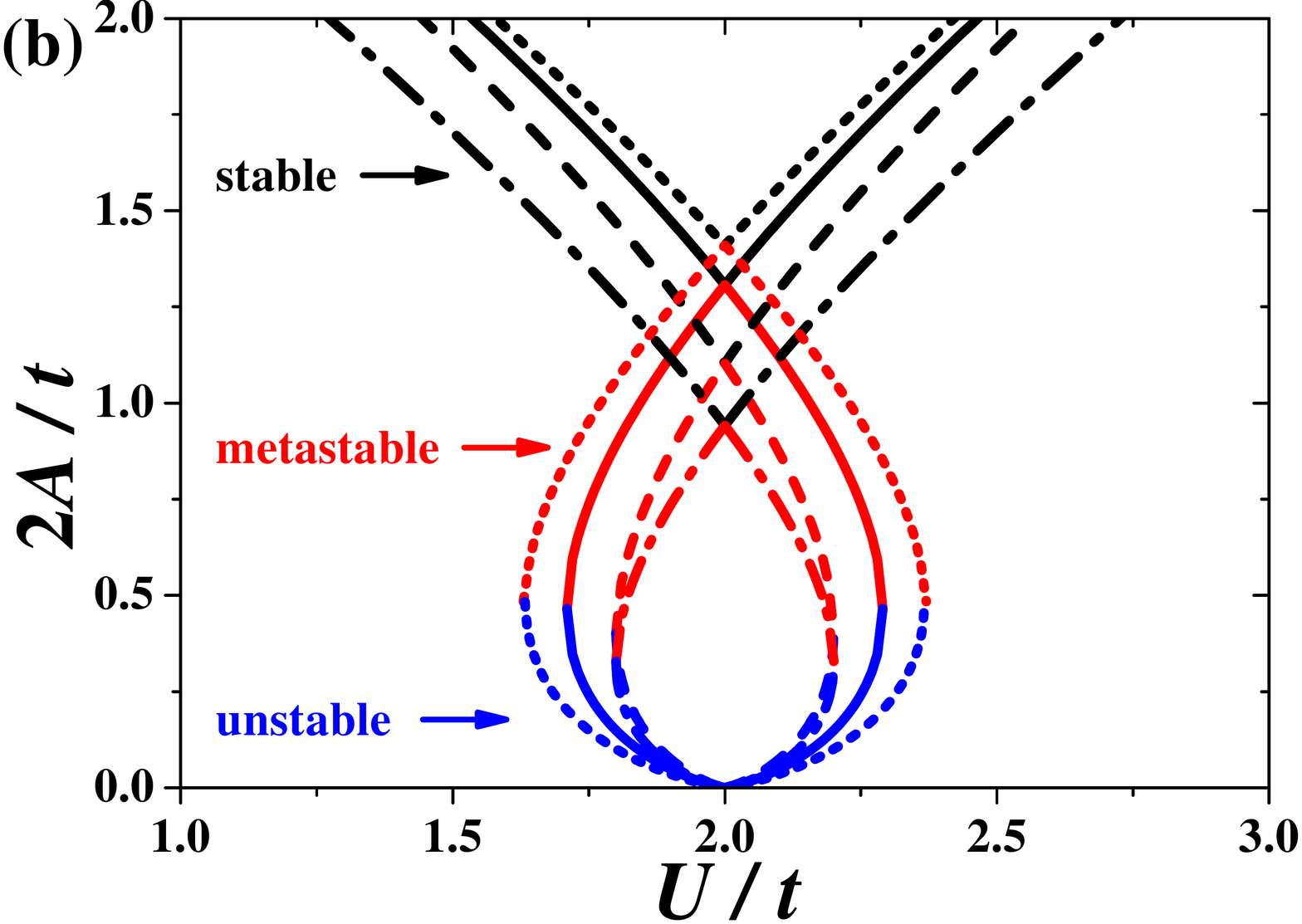}
   \includegraphics[width=\sizetwoTHREE]{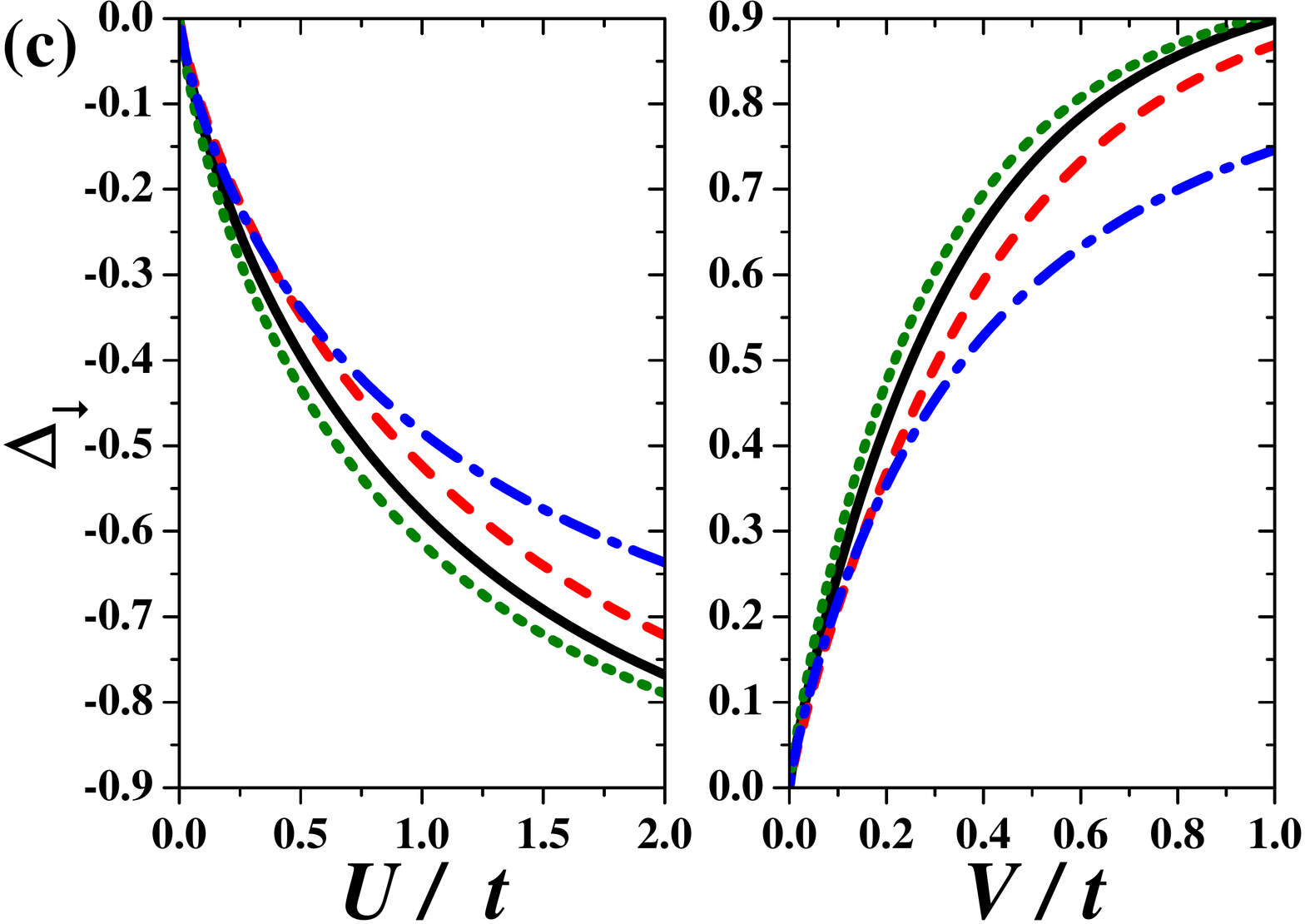}
	\caption{The dependence of $\Delta_{\downarrow}$ [panels (a) and (c)] 
	and energy gap $2A/t$ [panel (b)] as a function of model parameters for different DOSs at $T=0$.
	The solid, dotted, dot-dashed, and dashed lines correspond to 
	the semi-elliptic, gaussian, lorentzian, and rectangular DOSs, respectively 
	(all results obtained within the HFA).
	On panels (a) and (b), black, red, and blue color correspond to 
	stable, metastable, and unstable solutions, respectively.
	The results are obtained for $V/t=1.0$ [panels (a) and (b)], 
	$V/t=0.0$ [panel (c), left], and $U/t=0.0$ [panel (c), right].}
	\label{fig:GSproperties}
\end{figure*}%

The diagram of the EFKM for $T>0$ is determined by finding all solutions of the set of Eqs. (\ref{eq:DeltaDown})--(\ref{eq:DeltaUp}), comparing their free energies and checking if they correspond to the local minima of free energy $F$ [Eq. (\ref{eq:freeener})] with respect to $\Delta_{\downarrow}$ and $\Delta_{\uparrow}$ parameters.
The set usually has one solution (we restricted ourselves to the case of $\Delta_{\uparrow} \geq 0$) corresponding to the stable phase (i.e., free energy $F$ has a single minimum).
Only in some restricted ranges it has two solutions.
In such a case, $F$ has two local minima and the minimum with lower (higher) free energy corresponds to a stable (metastable) phase.
The parameters $\Delta_{\downarrow}$ and $\Delta_{\uparrow}$ characterize a phase occurring in the system.

\begin{figure*}	
   \includegraphics[width=\sizethree]{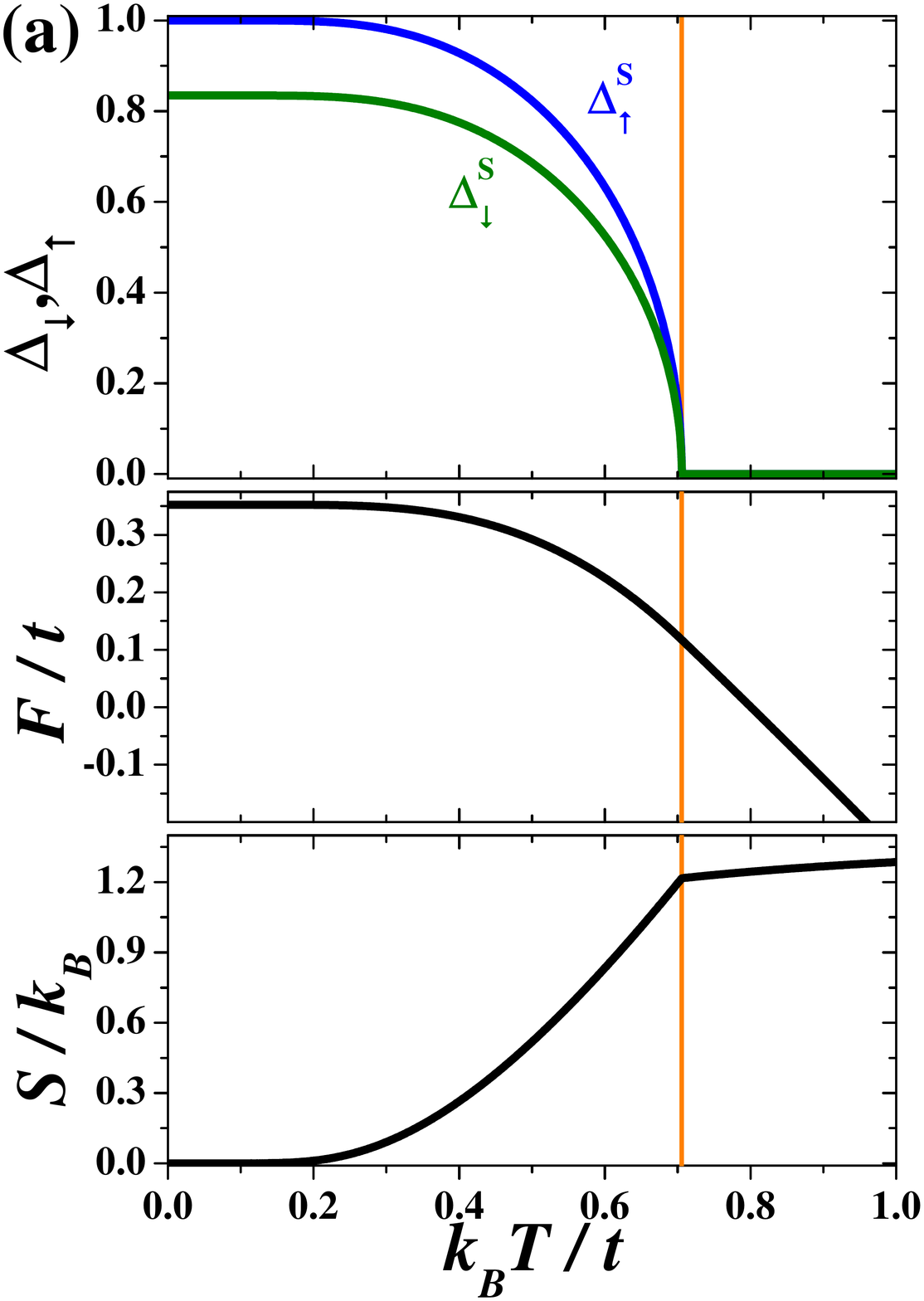}
   \includegraphics[width=\sizethree]{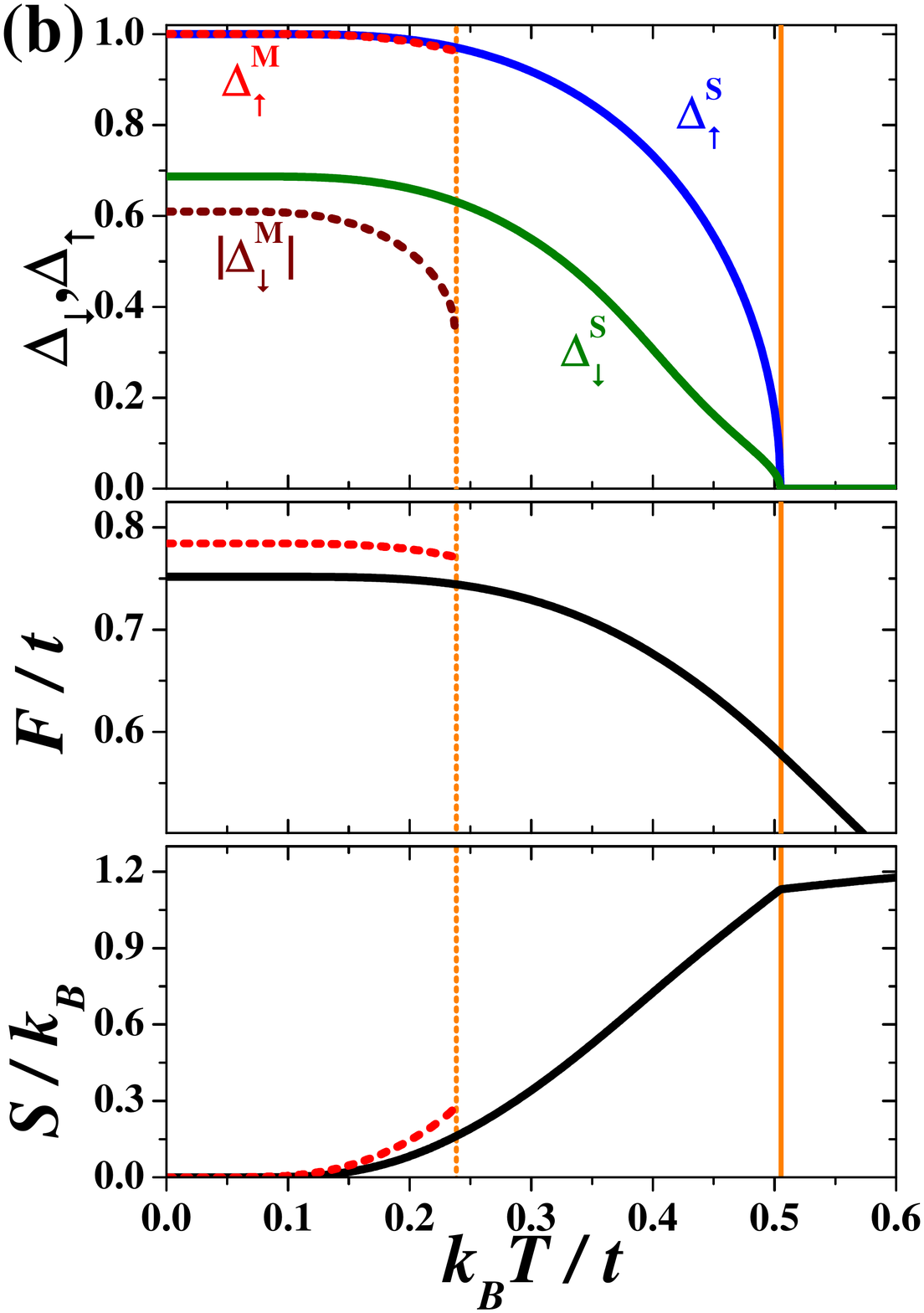}
   \includegraphics[width=\sizethree]{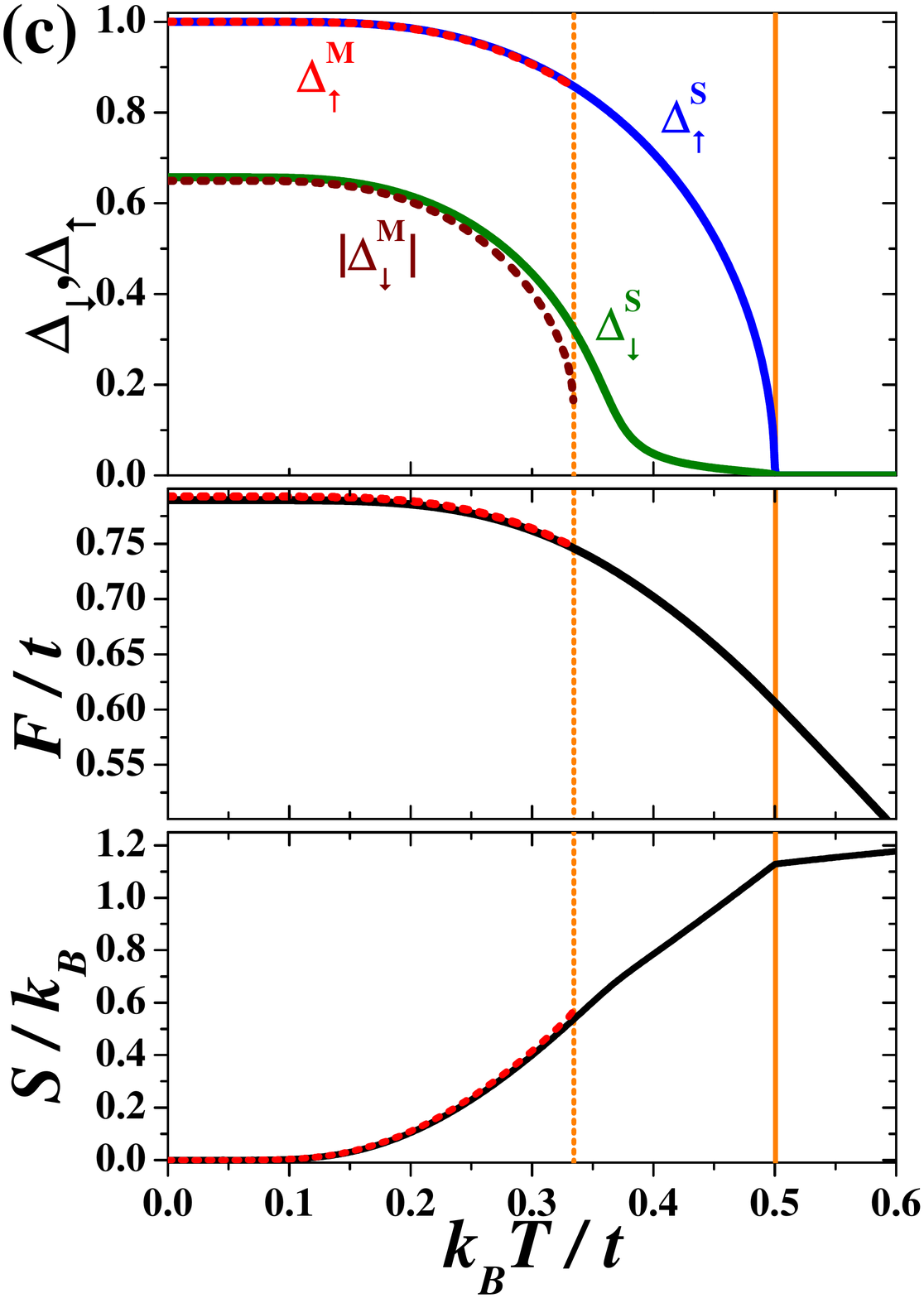}
	\caption{
		Temperature dependencies of parameters $\Delta_{\uparrow}$ and $\Delta_{\downarrow}$,
		free energy $F/t$ and  entropy $S/k_{B}$ (from the top to the bottom row) 
		for $V/t=1.0$ and: (a) $U/t=1.00$, (b) $U/t=1.90$, and (c) $U/t=1.99$.
		The solid and dotted lines correspond to stable and metastable solutions 
		('$S$' and '$M$' indexes, respectively).
		Vertical solid and dotted lines indicate temperature $T_{c}$ of the 
		continuous order-disorder transition and temperature, at which the metastable solution vanishes.
	}
	\label{fig:HFAthermodynamicquantitiesone}
\end{figure*}

The general structure of the finite temperature phase diagram of the model obtained within the HFA for fixed $V/t$ is not dependent on particular value of $V/t \neq 0$.
The exemplary phase diagram for $V/t=1.0$ is shown in Fig. \ref{fig:HFAexphasediagram}(a) [for other values of $V/t$ see also Fig. \ref{fig:DMFTvsHFAphasediagrams}].
It consists of three regions.
For large temperatures the NO phase is stable. 
With decreasing temperature, the continuous order-disorder transition at $T_c$ occurs [which coincides with a solution of (\ref{eq:crittemp})].
For $U<2V$, the transition is to the CO phase (i.e., the phase, where charge-order dominates over antiferromagnetic order), whereas, for $U>2V$, the low-temperature phase is the AF phase (antiferromagnetism dominates). 
$T_c$ is minimal for $U=2V$ and it is equal to $k_{B}T_{c}=V/2$.
It increases with increasing $|U-2V|$.
For $U=2V$ and $T<T_{c}^{*}$ [$k_{B}T_{c}^{*}/t = 0.357t$ for $V/t=1.0$; $\beta_{c}^{*} = (k_{B} T_{c}^{*})^{-1}$ is a solution of equation $I_{c}(\beta_{c}^{*}) = 1/V$, cf. Appendix \ref{sec:appU2Vtransition}], a discontinuous (first order) transition occurs between two different ordered phase.
The discontinuous boundary starts at $T=0$ and ends at isolated critical point (labeled by $I$-point) for $T=T_{c}^{*}$ (similarly as other found, e.g., for atomic limit of the model \cite{MicnasPRB1984,KapciaPhysA2016}). 
In Fig. \ref{fig:HFAexphasediagram}, also regions of occurrence of the metastable phases (near discontinuous transition) are determined. 
A range of $U/t$ where both phases (i.e., the CO and AF phases) coexist is  $1.710<U/t<2.290$ at the ground state and vanishes at $T=T_{c}^{*}$.
For $U=2V$ and $T_{c}^{*}<T<T_{c}$, there is no transition but only  a smooth crossover between the CO and AF phases occurs through the point, where both antiferromagnetic and charge order parameters are the same, and none of them is dominant (i.e., $\Delta_Q = m_Q$).
Although, at $U=2V$, the parameter $\Delta_{\downarrow}=0$, but $\Delta_{\uparrow}=\Delta_Q = m_Q>0$, thus the system is still in the ordered phase at $T_{c}^{*}<T<T_{c}$ (cf. also  Fig. \ref{fig:HFAthermodynamicquantitiesone} and Appendix \ref{sec:appU2Vtransition}). 
Please note that, in Fig.~\ref{fig:HFAexphasediagram}(a), the order-disorder transition line (which can be continuous as well as discontinuous dependently on $U/t$) and the discontinuous CO--AF boundary obtained within the DMFT are also shown by grey lines. 
They are taken from Ref. \cite{KapciaPRB2019}.
The lines of metal-insulator transitions are not indicated there.

In Fig. \ref{fig:HFAexphasediagram}(b), we present a dependence of order-disorder transition temperature $T_c$ as a function of $\left( U - 2V \right) /t$ for different  values of $V/t$.
One can notice that it is an increasing function of $|U-2V|$ with the minimal value of $k_{B}T_{c}(U=2V) = V/2$ [this results is easily obtained from Eq. (\ref{eq:crittemp})].
With increasing $V$ the lines $k_{B}T_{c}-V/2$ (as a function of $U-2V$) line are one above the other [i.e., $T_{c}(U-2V,V=V_1) \leq T_{c}(U-2V,V=V_2)$ if $V_1<V_2$].      
For $V/t \rightarrow + \infty$ or $U\rightarrow \pm \infty $ (which is equivalent with $t \rightarrow 0$) the critical temperatures approaches $k_{B}T_{c} \rightarrow V/2 + | U - 2V | / 4$ (cf. Sec. \ref{sec:equationTc}).
In particular, for $U=0$, $T_{c}$ approaches the results for the atomic limit of the model: $k_{B}T_{c}(U=0)=V$ \cite{MicnasPRB1984,KapciaPhysA2016} (obviously, for $U \neq 0 $ it does not coincide with rigorous results for the $t=0$ limit).
One should also stress that the first order boundary is located at $U=2V$ for any $V/t$.

Figure \ref{fig:HFATcdifferentDOS}(a) shows $U$-dependence of $T_c$ for $V/t=1.0$ and different DOSs (listed in Sec. \ref{sec:modelandmethod}).
One can see that  $k_{B} T_{c} (U=2V)=V/2$ independently on the DOS used for calculations (Appendix \ref{sec:appU2Vtransition}).
Moreover, there is no significant qualitative difference between all the curves, however, they are different quantitatively.  
Namely, temperatures $T_c$ obtained for gaussian $D_G(\varepsilon)$ are slightly higher that those obtained for semi-elliptic $D_{S-E}(\varepsilon)$. 
The line of $T_{c}$ for $D_{R}(\varepsilon)$ is located below the curve obtained for the semi-elliptical DOS.
The lowest critical temperatures $T_{c}$ are calculated for the lorentzian DOS.
This sequence of $T_{c}$ curves occurs for any $V/t$.
One should notice that, for $V\rightarrow + \infty$ or $U\rightarrow \pm \infty$, one gets $k_{B}T_{c} = V/2 + |V/2 - U/4|$ (for any symmetric DOS this is proven analytically  in Sec. \ref{sec:equationTc}), but this behavior is not visible in Fig. \ref{fig:HFATcdifferentDOS}(a), which is obtained for relatively small $V/t$ and $U/t$.
Note also that, for $U=2V$, the first order boundary for temperatures smaller than $T_{c}^{*}$ (which depends on the DOS) occurs as well as the smooth crossover region for $T_{c}^{*}<T<T_{c}$ (not shown in the figure).
The dependence of $T_{c}^{*}$ temperature (which is located for $U=2V$) as a function of $V/t$ and different DOSs is shown in Fig. \ref{fig:HFATcdifferentDOS}(b).
It increases monotonously with increasing  $V/t$ from $T_{c}^{*}/V=0$ (at $V/t\rightarrow 0$) to $k_{B}T_{c}^{*}/V=k_{B} T_{c} /V \rightarrow 0.5$ (at $V/t\rightarrow \infty$) for any DOS. 
However, one should note that the lines  $T_{c}^{*}/V$ obtained for the lorentzian and rectangular DOSs cross at $V/t\approx 0.632$ and this obtained for $D_{L}(\varepsilon)$ is above that calculated for $D_{R}(\varepsilon)$.

\subsubsection*{Changes of thermodynamic quantities at phase boundaries}

The ground state properties of the model for the Bethe lattice were studied in detail in Ref. \cite{LemanskiPRB2017}.
Here, we only present the behavior of $\Delta_{\downarrow}$ and energy gap $2A/t$ in the neighborhood of the discontinuous transition at $U=2V$ for different DOSs as a function of $U/t$ (for fixed $V/t=1.0$), cf. also Fig. 6 of Ref. \cite{LemanskiPRB2017}.
They are shown in Fig. \ref{fig:GSproperties}(a) and (b).
The obtained results shows that in stable and metastable phases $\Delta_{\downarrow}$ and the energy gap at the Fermi level for different DOS are in the same order as $T_{c}$, i.e., 
the biggest one is that for $D_{G}(\varepsilon)$, next are those for  $D_{S-E}(\varepsilon)$ and $D_{R}(\varepsilon)$, and the smallest is that for $D_{L}(\varepsilon)$. 
In the unstable solution the order is inverted.
Also the range of the coexistence region is dependent on the used DOS (the biggest for $D_{G}(\varepsilon)$, the smallest for $D_{L}(\varepsilon)$).  
We also presented $\Delta_{\downarrow}$ as a function of $U/t$ [for $V/t=0$, the left panel of Fig. \ref{fig:GSproperties}(c)] as well as as a function of $V/t$ [for $U/t=0$, the right panel of Fig. \ref{fig:GSproperties}(c)].
For large $U/t$ and $V/t$ parameter $|\Delta_{\downarrow}|$ goes to $1$ for any DOS. 
Please note that for small values of $U/t$ and $V/t$ the lines obtained for $D_{R}(\varepsilon)$ and $D_{L}(\varepsilon)$ cross each other.

\begin{figure*}	
   \includegraphics[width=\sizethree]{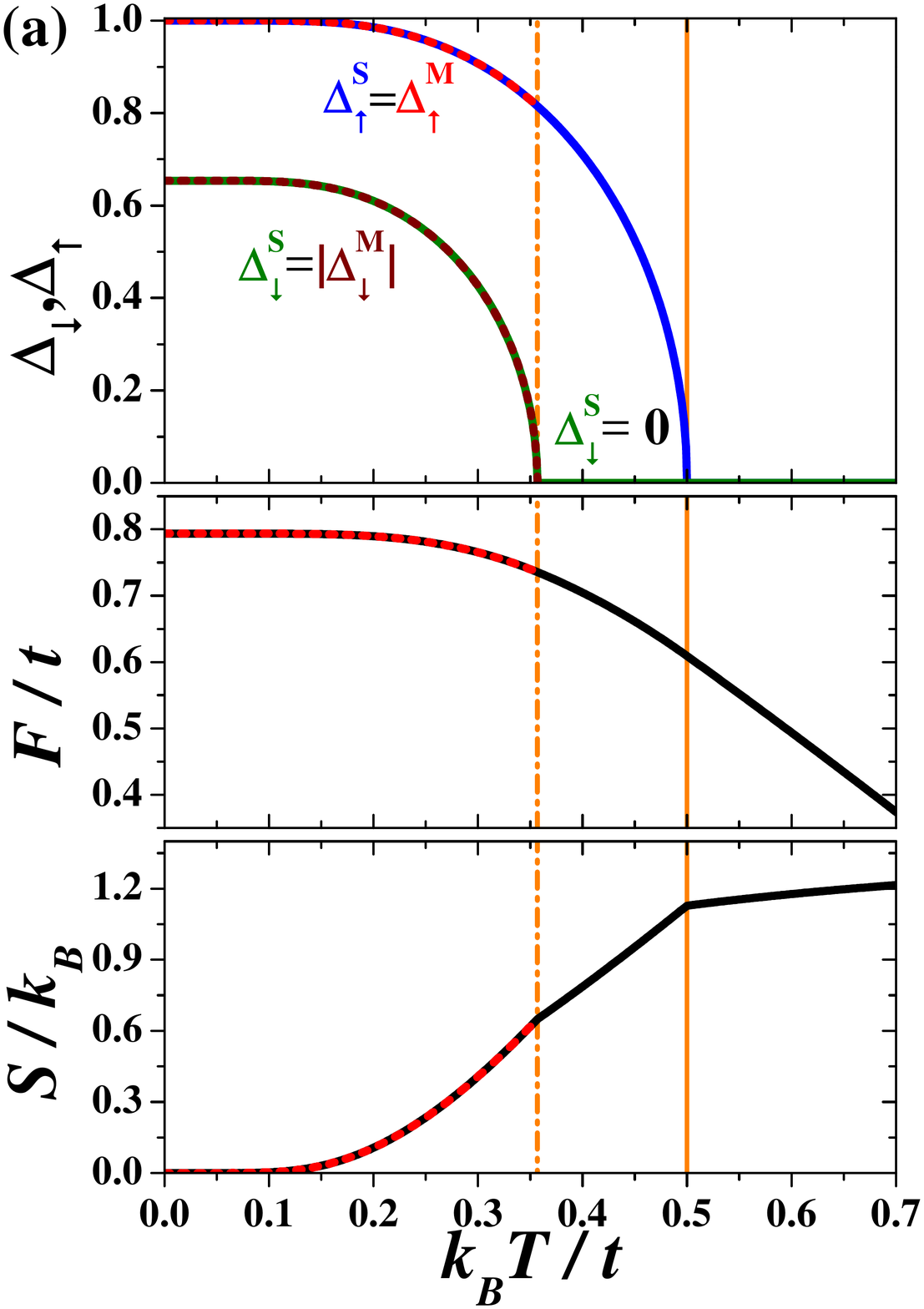}
   \includegraphics[width=\sizethree]{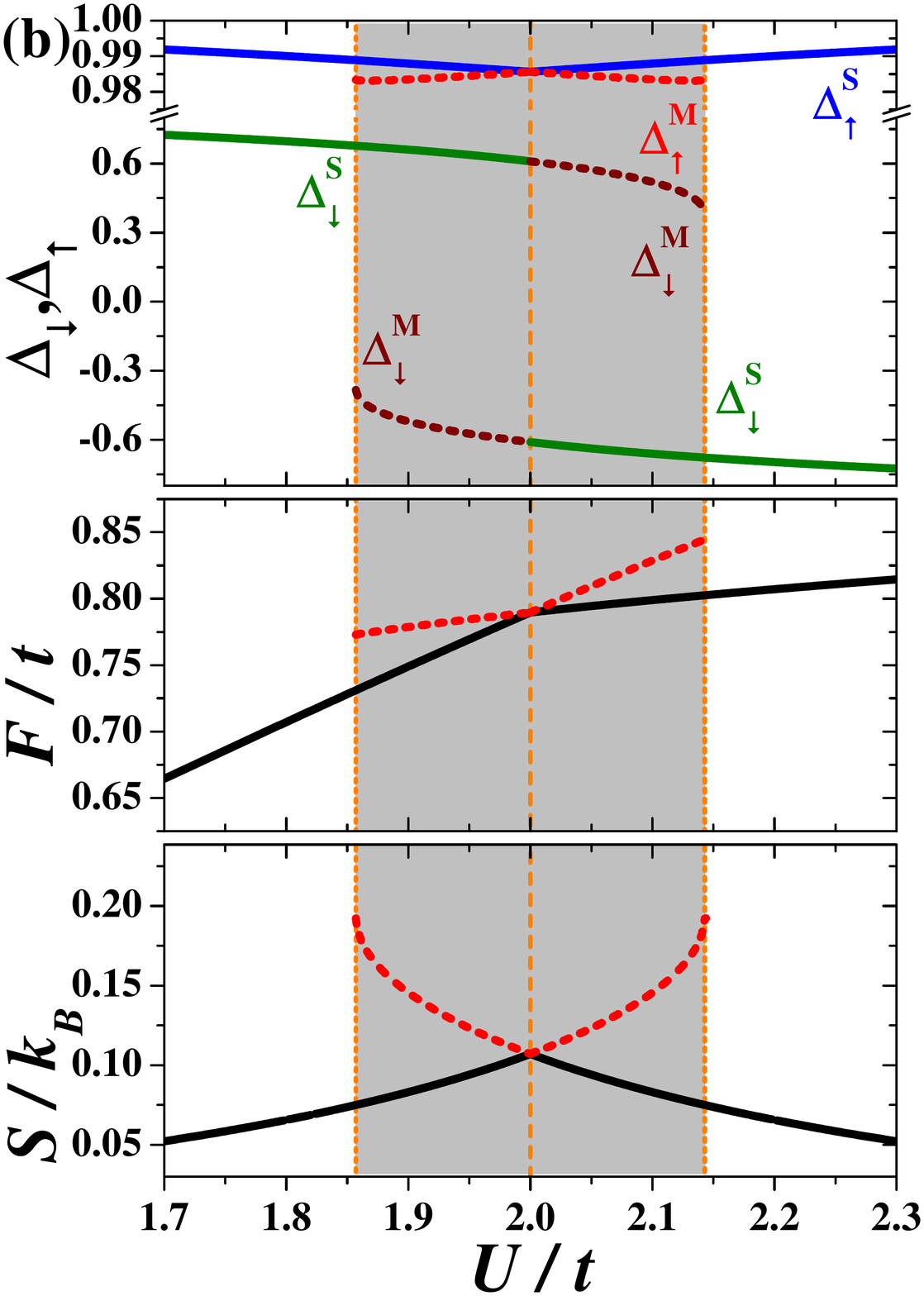}
   \includegraphics[width=\sizethree]{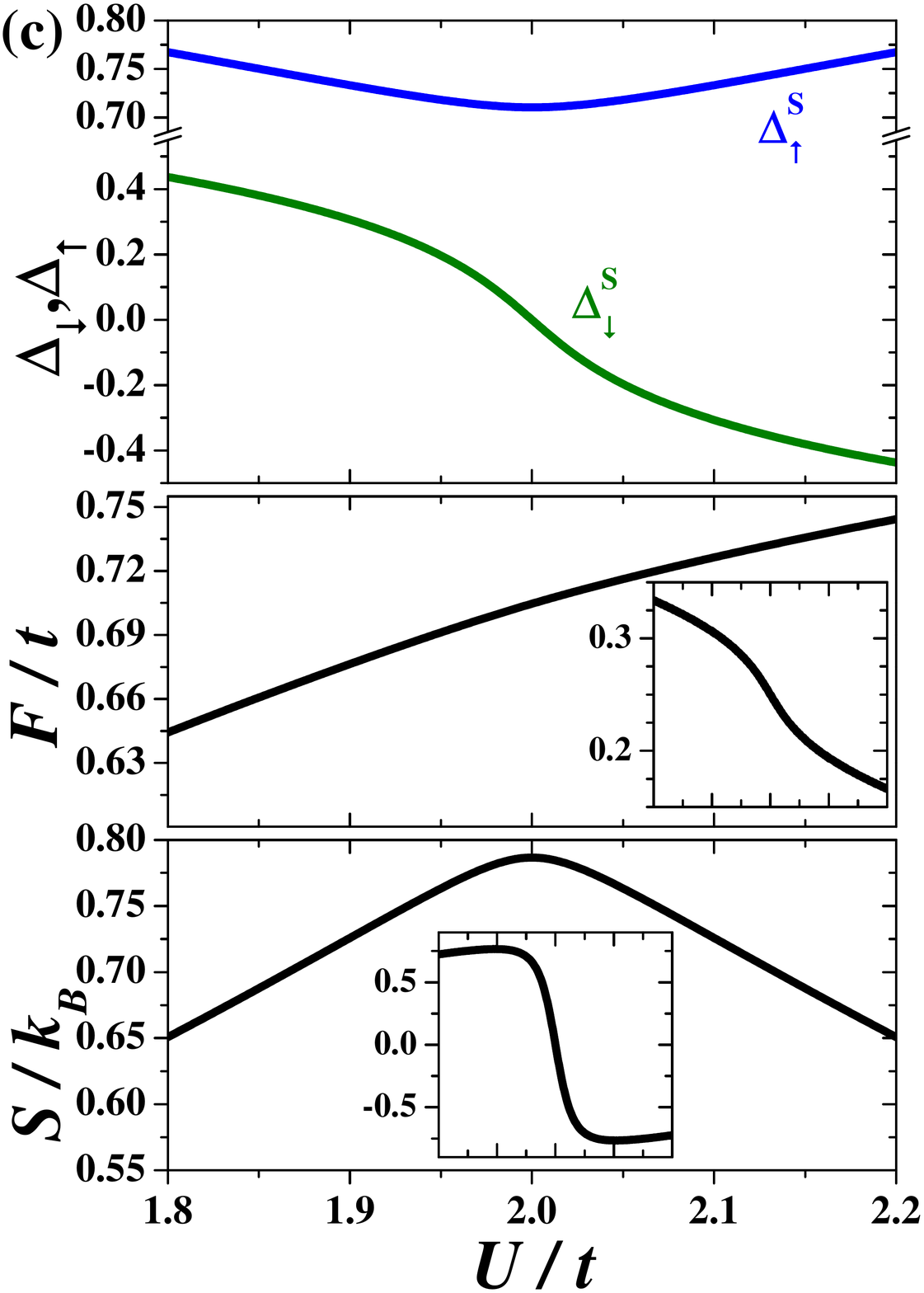}
	\caption{
		(a) Temperature dependencies of parameters $\Delta_{\uparrow}$ and $\Delta_{\downarrow}$,
		free energy $F/t$ and  entropy $S/k_{B}$ (from the top to the bottom row) 
		for $V/t=1.0$ and $U/t=2.0$.
		The vertical dash-dotted and solid lines indicates temperatures $T_{c}^{*}$ and $T_{c}$, respectively.
		Other denotations as in Fig. \ref{fig:HFAthermodynamicquantitiesone}.
		Note that, below $T_{c}^{*}$, 
		both stable and metastable solutions have the same $\Delta_{\uparrow}$, $F$, and $S$. 		
		(b), (c) The same thermodynamic parameters as a function of $U/t$ 
		for $V/t=1.0$ and $k_{B}T/t=0.2$, i.e, $T<T_{c}^{*}$ [panel (b)] 
		or  $k_{B}T/t=0.4$, i.e, $T>T_{c}^{*}$ [panel (c)]. 
		In the small insets $U$-dependence of derivatives 
		$\partial F / \partial U$ and $\partial S / \partial U$ are shown 
		(the second one in the units of $k_{B}/t$, the range of the horizontal scale is the same as that for the main figure). 
		The dotted lines correspond to the metastable solutions.
		The light grey shadow and the  vertical dashed lines indicate
		the coexistence region and the CO--AF discontinuous transition, respectively.   
		}
	\label{fig:HFAthermodynamicquantitiestwo}
\end{figure*}

\begin{figure*}
	\includegraphics[width=\sizethree]{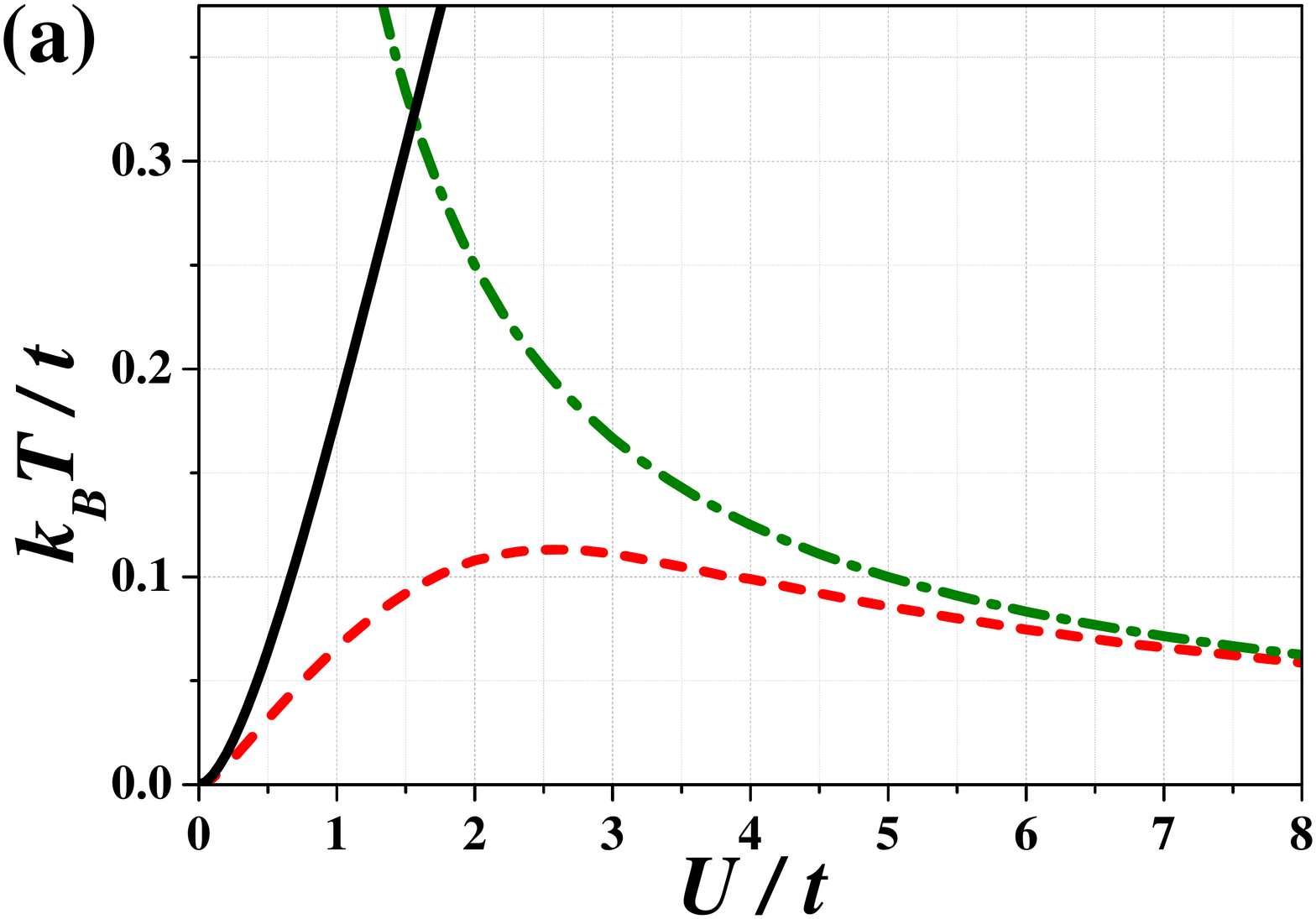}
	\includegraphics[width=\sizethree]{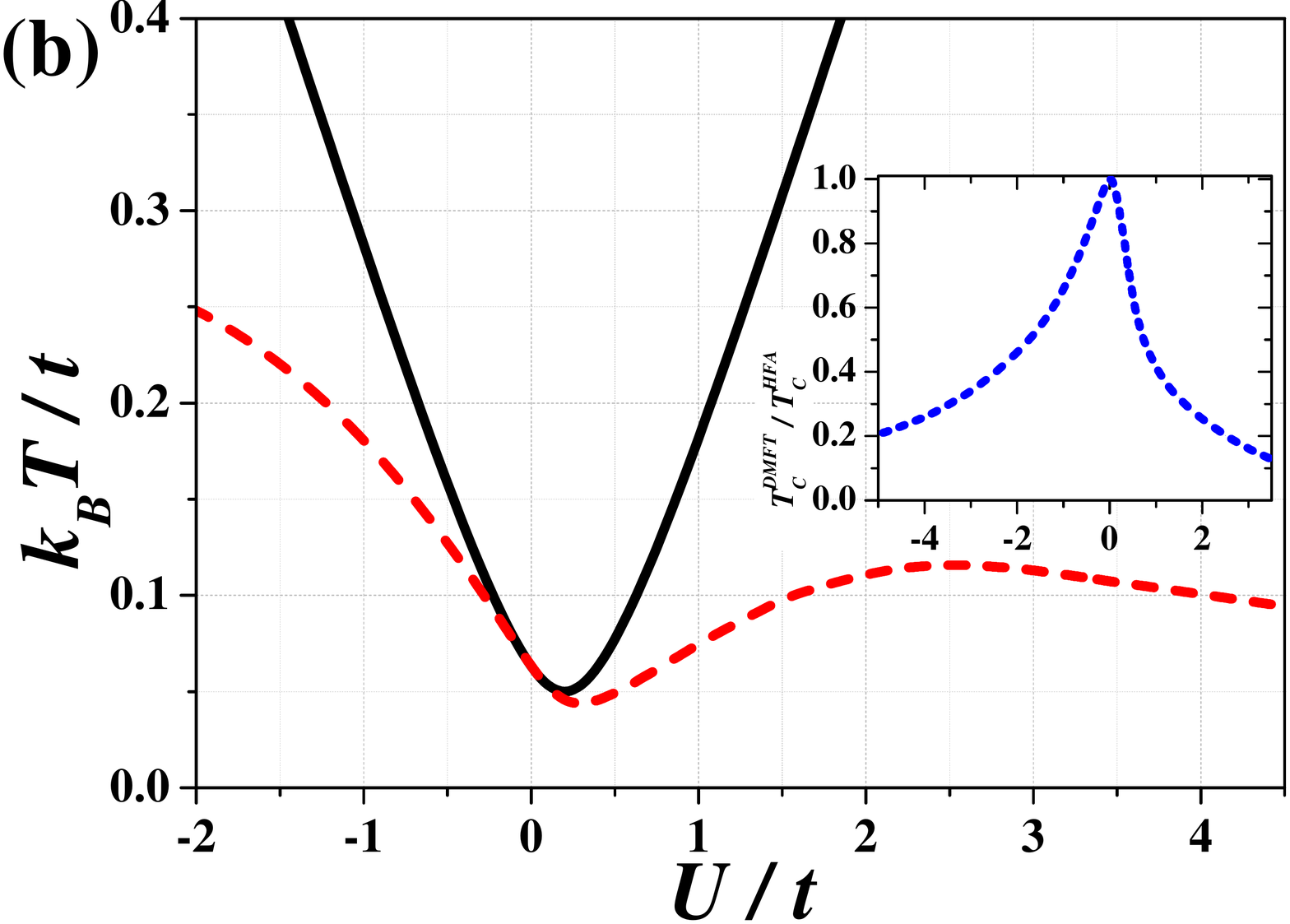}
	\includegraphics[width=\sizethree]{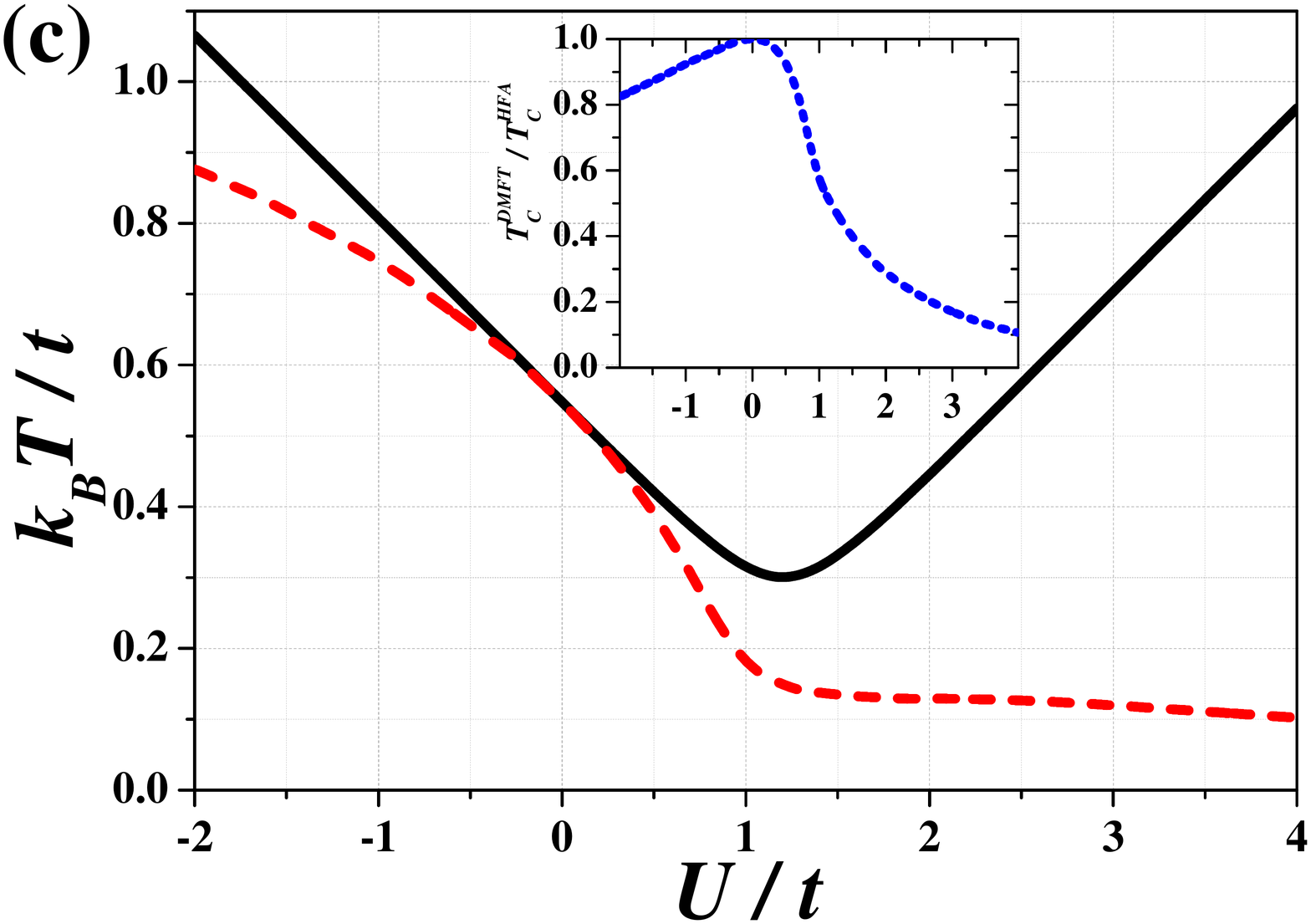}
	\caption{
		The phase diagrams of the model for $V/t=0.0$ (a), $V/t=0.1$ (b), and $V/t=0.6$ (c).
		The solid  and dashed  lines denote the continuous order-disorder 
		transition calculated within the HFA and the DMFT, respectively.
		The other transitions occurring on the diagram are not marked here. 
		In the insets of panels (b) and (c), the ratio $T_{c}^{DMFT}/T_{c}^{HFA}$ 
		of these two temperatures is also shown (the dotted lines).
		The dash-dotted green line on panel (a) denotes $t/(2U)$ line. 
		All results obtained for the semi-elliptic DOS.
		The DMFT results are taken from Refs. \cite{Lemanski2014,KapciaPRB2019}.
	}
	\label{fig:DMFTvsHFAphasediagrams}
\end{figure*}

In Figs. \ref{fig:HFAthermodynamicquantitiesone} and \ref{fig:HFAthermodynamicquantitiestwo}, a few representative dependencies of thermodynamic quantities are shown as a function of temperature or onsite interaction $U/t$ for $V/t=1$.
Apart from the quantities defined in Sec. \ref{sec:modelandmethod}, the behavior of entropy per site, defined as $S=-\partial F / \partial T$, is also shown.

Figure \ref{fig:HFAthermodynamicquantitiesone}(a) presents the behavior of parameters $\Delta_{\downarrow}$, $\Delta_{\uparrow}$, free energy $F$, and entropy $S$ for the region where any metastable phase does not exist.
Below $T_{c}$ the CO phase is stable (with both $\Delta_{\uparrow}>0$ and $\Delta_{\downarrow}>0$), whereas for $T>T_{c}$ the NO phase occurs.
Parameters $\Delta_{\downarrow}$ and $\Delta_{\uparrow}$ exhibit standard mean field dependencies and vanish continuously at $T_{c}$ as expected for the continuous (second order) transition.
$F$ and $S$ are continuous at the transition temperature. 
It is clearly  seen that the slope of $S$ [associated with a specific heat $c=(1/T) \partial S / \partial T = - (1/T) \partial^2 F / \partial T^2 $] exhibits a discontinuity at $T_{c}$.

\begin{figure*}	
   \includegraphics[width=\sizetwobig]{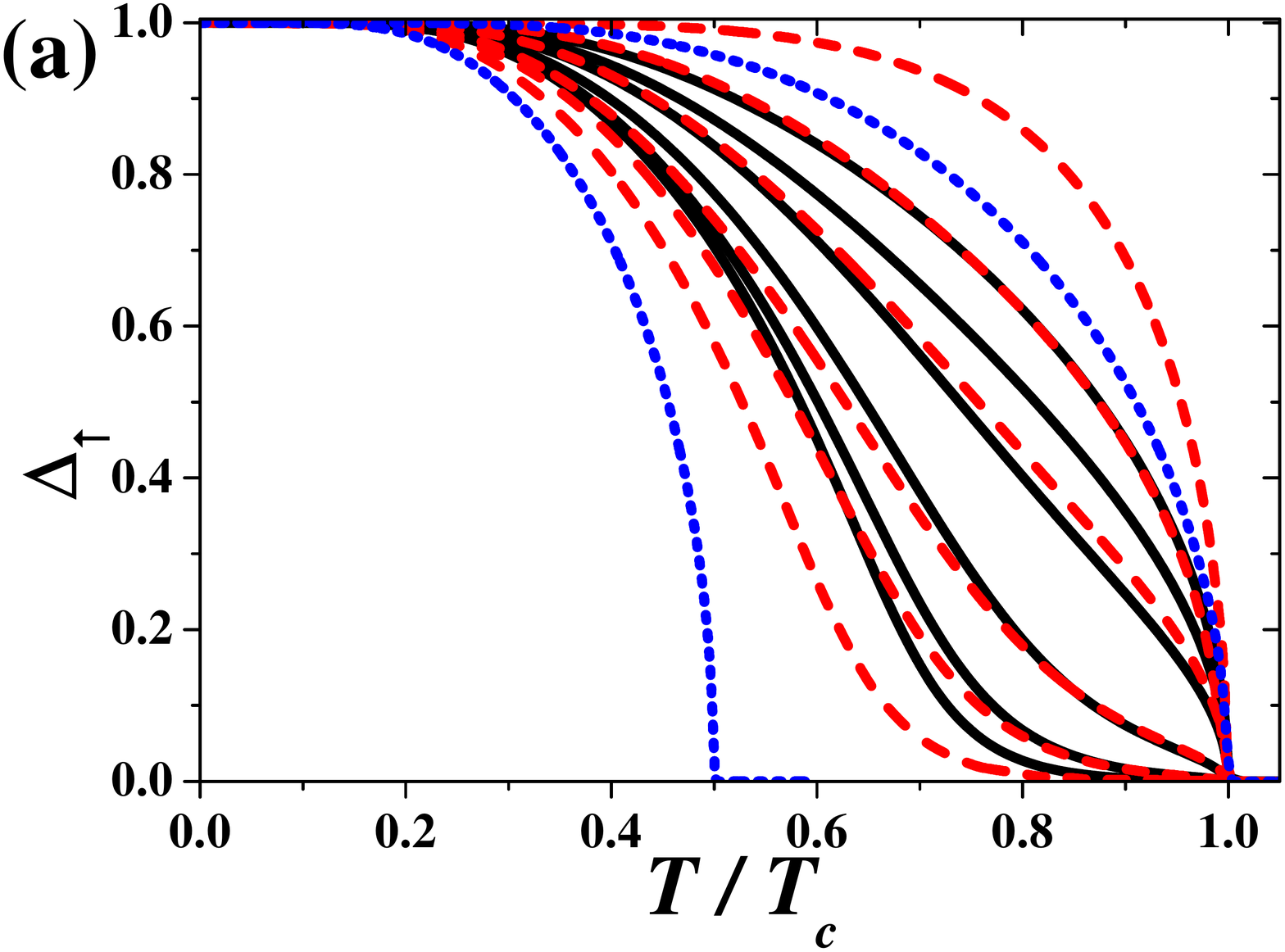}
   \includegraphics[width=\sizetwobig]{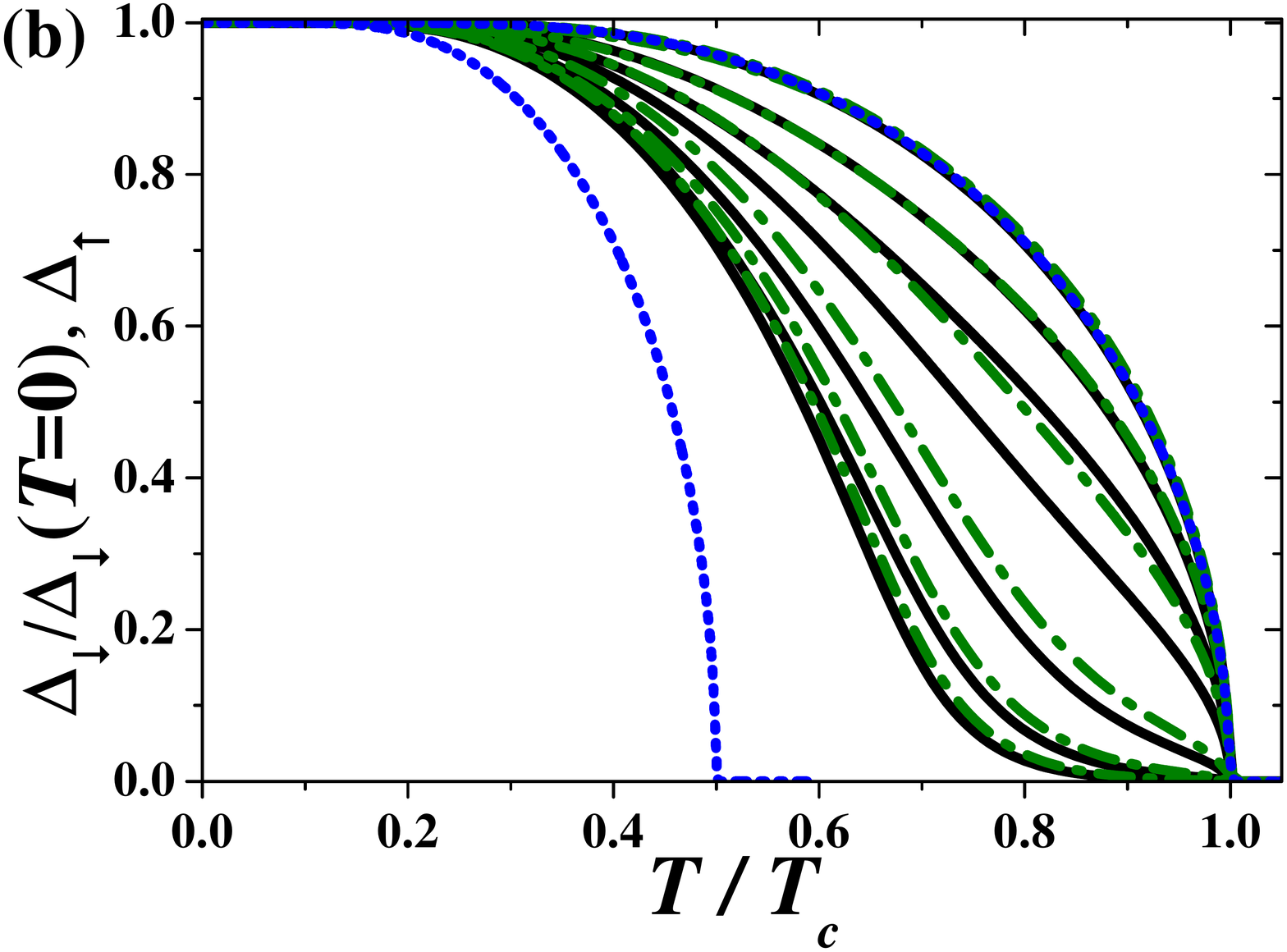}
	\caption{(a) Comparison between dependencies of $\Delta_{\uparrow}$ obtained within the HFA (solid black lines) and the DMFT (taken from Ref. \cite{Krawczyk2018}, dashed red lines) for $V=0$ and small values of $U/t$ ($U/t=0.0001$, $0.001$, $0.01$, $0.1$, $0.3$, $1.0$, from the bottom to the top, respectively).
	(b) Dependencies of normalized $\Delta_{\downarrow}/\Delta_{\downarrow}(T=0)$ (dash-dotted green lines) and $\Delta_{\uparrow}$ (solid black lines )  for $V=0$ and $U/t=0.0001$, $0.001$, $0.01$, $0.1$, $0.3$, $1.0$, $10.0$ (from the bottom to the top) obtained within the HFA.
		The results are obtained for the semi-elliptic DOS.
		The dashed blue lines denote the standard dependence of the mean-field order 
		parameter with transition at $T=T_{c1}^{Is}=T_{c}$ and with $T=T_{c2}^{Is}=T_{c}/2$ 
		[i.e., the solutions of Eq. (\ref{eq:magnetIsing})]. 
	}
	\label{fig:dsmallUHFAvsDMFT}
\end{figure*}

In Figs. \ref{fig:HFAthermodynamicquantitiesone}(b) and \ref{fig:HFAthermodynamicquantitiesone}(c) the temperature dependencies of the parameters are shown for such values of $U/t$ that metastable solutions exist in low temperatures.
The stable CO phase is characterized by $\Delta_{\downarrow}>0$, whereas in the metastable AF phase $\Delta_{\downarrow}<0$ (they absolute values differ slightly, the smaller one is those in the metastable phase).
The value of $\Delta_{\uparrow}$ is also barely smaller in the AF phase than that in the stable CO phase. 
These differences decrease with approaching $U=2V$ ($U=2.0$ in this particular example).
As one can expect, free energy $F$ and entropy $S$ take higher values in the metastable AF phase (however, the differences are relatively tiny, particularly for $U=1.99$).   
The CO--NO transition at $T_{c}$ is continuous, but the temperature dependency of $\Delta_{\downarrow}$ (in the stable phase) for intermediate temperatures below $T_{c}$ is deflected from standard mean field dependence.
However, function $\Delta_{\downarrow}(T)$ has still the square root character when it approaches $T_c$. 
This deflection is associated with the fact that for $U=2V$, $\Delta_{\downarrow}$ vanishes continuously at $T_{c}^{*}$, which differs from $T_{c}$ [cf. Fig. \ref{fig:HFAthermodynamicquantitiestwo}(a)].
Note that, in such a case, the CO and AF phases with, respectively, $\Delta_{\downarrow} \geq 0$ and $\Delta_{\downarrow}^{'}=-\Delta_{\downarrow} \leq 0$ have exactly the same energy $F$ and entropy $S$ (the solution for the AF phase is also shown in the figure by dotted lines).
For extremely small $U/t$ and $V/t$ one observes the huge deviation from standard square-root temperature behavior for  $\Delta_{\downarrow}$ and $\Delta_{\uparrow}$ parameters, but temperature dependence of normalized $\Delta_{\downarrow}/\Delta_{\downarrow}(T=0)$ almost follows behavior of $\Delta_{\uparrow}$ for $V/t=0$.
Some discussion concerning temperature dependence of these quantities for small values of interaction $U$ is also presented in Sec.~\ref{sec:validityHFA} and Fig.~\ref{fig:dsmallUHFAvsDMFT}.

The analysis of behavior of the system for $U=2V$ is presented in Fig. \ref{fig:HFAthermodynamicquantitiestwo}.
For small temperatures $T<T_{c}^{*}$ [Fig. \ref{fig:HFAthermodynamicquantitiestwo}(b)], the CO--AF transition with changing $U/t$ exhibits typical behavior observed for a discontinuous (first order) transition.
One can identify the coexistence region in the neighborhood of the boundary.
For $U<2V$, the CO phase with $\Delta_{\downarrow}>0$ is stable and the AF phase  with $\Delta_{\downarrow}<0$ is metastable (it means that the free energy of the AF phase is higher than that of the CO phase), whereas, for $U>2V$, the situations is opposite, namely the AF phase is stable and the CO phase is metastable.
$\Delta_{\downarrow}$ exhibits a discontinuous jump at the transition and never takes the value $0$, even though the parameter $\Delta_{\uparrow}>0$ changes continuously.
It is also clearly seen that slopes of $F$ and $S$ with respect to $U$ changes discontinuously at $U=2V$ transition point (for $T<T_{c}^{*}$).
For $T_{c}^{*}<T<T_{c}$ the parameter $\Delta_{\downarrow}$ as a function of $U/t$ changes its sign continuously going through $0$-value (with simultaneous continuous behavior of $\Delta_{\uparrow}>0$ as well as $n_Q$ and $m_Q$).
Note also that one cannot find any indicators of a phase transition in the $U$-dependence of $F$ and $S$.
Their slopes are continuous functions for $U=2V$ (in the insets $\partial F/ \partial U$ and $\partial S/ \partial U$ 
are presented).
Thus, at $T_{c}^{*}$ and $U=2V$, the isolated-critical $I$-point is identified.
In this point, the first-order CO--AF boundary ends inside the region of the ordered phase occurrence.
Above $T_{c}^{*}$ both phases are not distinguishable thermodynamically (with $\Delta_{\downarrow}=0$ and $m_Q=\Delta_Q=\Delta_{\uparrow}$ at $U=2V$) and crossing of $U=2V$ is not associated with a phase transition.
For $U=2V$, the order-disorder transition occurs at $T=T_{c}$  (increasing temperature for fixed $U$) and it is associated with continuous vanishing of $\Delta_{\uparrow}$ [Fig.~\ref{fig:HFAthermodynamicquantitiestwo}(a)]. 
All thermodynamic quantities behave as for a standard second-order transition.
Note also that at $T_{c}^{*}$ and $U=2V$ the thermodynamic quantities in the both stable phases also exhibits standard behavior for a continuous transition.
Nevertheless, two solutions with different $\Delta_{\downarrow}$ of opposite values exist at this line below $T_{c}^{*}$. 
At $U=2V$ any derivatives of $F$ (with respect to $T$) are continuous, even for $T<T_{c}^{*}$, where it is a first order transition,  because the transition is not the temperature transition [cf. the insets in Fig. \ref{fig:HFAthermodynamicquantitiestwo}(c)].
The boundary can be crossed only by changing the interaction value and the first derivative of $F$ (as well as of $S$) with respect to $U$ is indeed discontinuous for $T<T_{c}^{*}$.

Note that the transition at $U=2V$ at $T_{c}^{*}$ resembles the transition in the ferromagnetic Ising model in the external magnetic field occurring at field $H=0$ (obtained within the mean field approximation), e.g., Refs. \cite{Brush1967,Vives1997}.
From the temperature dependence of the total magnetization $M$, i.e., solving the equation 
\begin{equation}
\label{eq:magnetIsing}
M=\tanh \left[ \left(\frac{T_{c}^{Is}}{T} \right) M \right], 
\end{equation}
one finds the continuous transition at $T_{c}^{Is}$ (both solution with $M>0$ and 
$M<0$ are equivalent). 
If one investigates $M$ as a function of $H$ for $T<T_{c}^{Is}$, one finds the discontinuous transition at $H=0$ between phases with $M>0$ (stable for $H \geq 0$ and metastable for $H<0$) and $M<0$ (having the lowest energy for $H \leq 0$ and being metastable for $H>0$).
Note that in the vicinity of $H=0$ both solutions with $M>0$ and $M<0$ exists.

For better understanding the behavior of the model for $T_{c}>T>T^{*}_{c}$ let us also mention an analogy taken from the theory of magnetism.
Namely, for $T> T_{c}^{Is}$, the magnetization of a paramagnet in the external magnetic field changes from $M>0$ (if $H>0$) to $M<0$ (if $H<0$) continuously through $M=0$ with a change of direction of the external field $H$.
But at $H=0$, we do not observe any phase transition and $M=0$ at  $T> T_{c}^{Is}$, even though we can distinguish two regions on the phase diagram for $H \neq 0 $.
Obviously, we cannot construct a simple analogy between the Ising model and the continuous phase transition occurring at $T_{c}$ in the EFKM.

\subsection{The validity of the Hartree-Fock approach}
\label{sec:validityHFA}

Let us now discuss limitation of the HFA and compare the results derived within this approach with the results obtained by the DMFT.
As we said previously, the HFA gives rigorous results at the ground state (Sec. \ref{sec:expressionsGS} and Appendix \ref{sec:appHFAvsDMFT}, cf. also Ref. \cite{LemanskiPRB2017}).
For finite temperatures the situations is more complex.
In Fig. \ref{fig:DMFTvsHFAphasediagrams}, one can see the comparison of the order-disorder transition temperature obtained within the HFA and the DMFT (for various $V/t$). 
Notice that, for values of $V/t$ larger than approximately $0.7$, the DMFT predicts a discontinuous order-disorder transition.   
This behavior is not captured within the HFA (cf. also Fig.~\ref{fig:HFAexphasediagram}).
Nevertheless, for small $U$ both approaches give very comparable results in finite temperatures.
For $U=0$, they give exactly the same results, because in this limit both approaches are equivalent \cite{MullerHartmannZPB1989,KapciaJSNM2020}.

Please also note, that the HFA results do not agree with the results of the DMFT, which predicts that CO--AF boundary is a standard first order transition (for $T\neq 0$) associated with the discontinuity of $\Delta_\uparrow$ \cite{KapciaPRB2019,KapciaPRB2020}.
As it is shown above, at $T_{c}>T>T^*_{c}$, the HFA finds a smooth crossover between these two ordered phases (cf. Sec. \ref{sec:phasediagHFA}).

As one can expect the HFA fails in the description of metal-insulator transition.
Within the HFA all ordered phases are insulators with the gap $\Delta_F(\varepsilon)=2A$ ($\Delta_F(\varepsilon)\neq0$ if $\Delta_{\uparrow} \neq 0$), whereas the DMFT finds several metallic phases with the long-range order \cite{Lemanski2014,KapciaPRB2019,KapciaPRB2020}.

Finally, let us discuss the dependence of $\Delta_{\uparrow}$ as a function of reduced temperature $T/T_{c}$ for really small interactions parameters.
It was found that for small $U/t\rightarrow 0^{+}$ and $V/t=0$ the temperature dependence of $\Delta_{\uparrow}$ is quite unusual, namely, it resembles mean-field solution of Eq. (\ref{eq:magnetIsing}), but with $T_{c}^{Is}=T_{c}/2$ \cite{DongenPRL1990,DongenPRB1992,ChenPRB2003,Krawczyk2018}.
In Fig. \ref{fig:dsmallUHFAvsDMFT}(a), we show the comparison of the DMFT results with those obtained within the HFA for small values of $U/t$ (for the Bethe lattice).
As one can see the HFA values are higher than those of the DMFT, but the similarity is noticeable.
In Fig. \ref{fig:dsmallUHFAvsDMFT}(b), we also presents the dependence of normalized $\Delta_{\downarrow}/\Delta_{\downarrow}(T=0)$ for $V/t=0$ and this quantity almost follows behavior of $\Delta_{\uparrow}$ ($\Delta_{\uparrow}(T=0)=1$ for any $U$ and $V$).

\section{Conclusions and final remarks}
\label{sec:conclusions}

In this paper, we studied the extended Falicov-Kimball model within the weak-coupling limit, i.e., using the mean-field broken-symmetry Hartree-Fock approach.
We determined the phase diagram of the model and showed some thermodynamic characteristics.
We found that the diagram is symmetric with respect to $U=2V$.
The order-disorder transition is continuous, whereas the transition line  between ordered phases at $U=2V$ is discontinuous and finishes at the critical-end point. 
The detailed analysis of this specific values of the model parameters was provided.
Many analytic derivations based on the HFA equation for the order parameters and the free energy were performed for different density of states.

We then compared the results obtained by the HFA with those derived using the DMFT for the Bethe lattice.
We showed that only at $T=0$ both these methods are equivalent in the whole range of coupling parameters. When the local interaction parameter $U=0$ then the same results are also obtained for $T>0$.
Moreover, we obtained quantitatively similar results when $T>0$ and $U$ is small.
However, we proved, that for not very small values of $U$ HFA completely fails at $T>0$, because it gives results that differ significantly in quantitative terms, or in some intervals of the parameters also qualitatively from those obtained using DMFT.

Thus, we showed systematically that properties of the correlated electron system derived on the basis of the static  (represented here by the HFA) and the dynamic  (represented by the DMFT) mean field theory for not too small $U$ and $T>0$ are significantly different. In particular, these differences are enhanced in the limit of strong correlations, when $U\rightarrow \infty $.
We expect that similar conclusions to the ones presented here also apply to other models of correlated electrons, although proving this can be much more complicated due to the difficulty of obtaining exact results for these models.

\begin{acknowledgments}
The authors express their sincere thanks to J. K. Freericks and M. M. Ma{\'s}ka for useful discussions on some issues raised in this work. 
K. J. K. acknowledges the support from the National Science Centre (NCN, Poland) under Grant SONATINA 1 no. UMO-2017/24/C/ST3/00276.
K. J. K. appreciates also  founding in the frame of a scholarship of the Minister of Science and Higher Education (Poland) for outstanding young scientists (2019 edition, no. 821/STYP/14/2019).
\end{acknowledgments}

\appendix

\section{Equivalence of the HFA and the DMFT for the EFKM at the ground state}
\label{sec:appHFAvsDMFT}

In Ref. \cite{LemanskiPRB2017} the exact analytic formulas for $\Delta_{\downarrow}$ and $F_0$ (for the EFKM on the Bethe lattice, i.e., for the semicircular DOS, and the half-filling) were presented (with $t=1$ treated as the unit of the energy).
They were derived by two different methods: in the HFA and within the DMFT.
In the first case, Eqs. (12) and (13) of Ref. \cite{LemanskiPRB2017} are in a coincidence with our results represented by Eqs. (\ref{eq:DeltadownGS}) and (\ref{eq:freeenergyGS}) and they are given by:
\begin{eqnarray}
\label{eq:HFAapp}
\Delta_{\downarrow}=\frac{A_{0}}{2\pi}\int \frac{\sqrt{K}}{\sqrt{W}}d\varepsilon,\  
F_{0}=E_{0} -\tfrac{1}{8\pi} \int{\sqrt{K W }d\varepsilon}, \qquad
\end{eqnarray} 
where $K \equiv K(\varepsilon)=4-\varepsilon^{2}$ and $W \equiv W(\varepsilon)= 4\varepsilon^2+A_{0}^{2}$.
The formulas obtained within the dynamical mean-field theory [Eqs. (8) and (9) of Ref. \cite{LemanskiPRB2017}] are given by
\begin{eqnarray}
\label{eq:DMFTapp}
\Delta_{\downarrow} = \frac{A_{0}}{4\pi} \int M d \xi, \  
F_{0} = E_{0} -\tfrac{1}{4\pi} \int \xi^2 M d \xi, \qquad 
\end{eqnarray}
where $M \equiv M(\xi)=\sqrt{(16+A_{0}^{2}-4\xi^{2})/(4\xi^2-A^{2}_{0})}$.

In Ref. \cite{LemanskiPRB2017} the question is posed, whether those formulas are the same (it was checked numerically with the accuracy error of the order of $10^{-50}$). 
As one can expect, the answer is positive for both equations. 
Indeed, all integrand functions are even, hence we can change the integration limits to the positive semiline \((0,+\infty)\), so all integrand variables will be positive.
The substitution $\xi^2=\varepsilon^2 + A_0^2/4$ leads to $d\xi = \varepsilon d \varepsilon / \xi = 2 \varepsilon d \varepsilon /\sqrt {W(\varepsilon)}$, $M(\xi) = \sqrt{K(\varepsilon)}/\varepsilon$, and $\xi^2 M(\xi) = W(\varepsilon) \sqrt{K(\varepsilon)} / (4 \varepsilon)$.
Finally, one gets:
\begin{eqnarray}
\label{eq:equalityHFADMFT}
\tfrac{1}{2}\int M d \xi = \int \frac{\sqrt{K}}{\sqrt{W}}d\varepsilon, \ 
\int \xi^2 M d \xi = \tfrac{1}{2}\int \sqrt{W K }d\varepsilon \qquad
\end{eqnarray}

It remains only to show that domains of integration coincide for both integrals. 
In the case of HFA expresions, the integrand is real and well-defined for $\varepsilon \in \left[0,2\right]$, whereas case of second DMFT integral $ \xi \in\left[A_0/2,\sqrt{16+A_{0}^{2}}/2\right]$. 
Putting these limits into formula for substitution shows that:
$\varepsilon=0$ maps onto $\xi=A_{0}/2$ and
$\varepsilon=2$ maps onto $\xi= \sqrt{16+A_{0}^{2}}/2$,
which proves that equalities (\ref{eq:equalityHFADMFT}) really holds.
Thus, expressions in Eqs. (\ref{eq:HFAapp}) and (\ref{eq:DMFTapp}) for $\Delta_{\downarrow}$ and $F_0$ are the same, respectively.
As a result, the equivalence of the DMFT and the HFA for the EFKM at the ground state (for the Bethe lattice and the half-filling) is rigorously proven.

\section{Analytic analysis of the boundary at the symmetric point of $U=2V$}
\label{sec:appU2Vtransition}

Let $\Delta_{\downarrow}$ and $\Delta_{\uparrow}$ be solutions of  (\ref{eq:DeltaDown})--(\ref{eq:DeltaUp}) for $U$, $V$ and $\beta$.
Under change $U \rightarrow U' = 4V - U$ and $\Delta_{\downarrow} \rightarrow \Delta^{'}_{\downarrow}=- \Delta_{\downarrow}$ one gets that $A \rightarrow A' = - U' \Delta_{\uparrow}/2 + V (\Delta_{\uparrow} + \Delta_{\downarrow}^{'}) = - A$ and
$B \rightarrow B' = - U' \Delta_{\downarrow}^{'} + 2V (\Delta_{\uparrow} + \Delta_{\downarrow}^{'}) = B$. 
Thus,  it is simply seen that $\Delta_{\downarrow}^{'}=- \Delta_{\downarrow}$ and $\Delta_{\uparrow}$ are also solutions of  (\ref{eq:DeltaDown})--(\ref{eq:DeltaUp}), but for $U'$, $V$ and $\beta$.

Notice also that the derivation presented here supports the results obtained numerically that the phase diagram of model (\ref{eq:ham}) derived within HFA needs to be symmetric with respect to $U=2V$ line.

From (\ref{eq:freeener}) and the relations derived above one also gets that 
$F(U,V,\Delta_{\downarrow}, \Delta_{\uparrow})=F(4V-U,V,-\Delta_{\downarrow}, \Delta_{\uparrow}) + (U-2V)/2$.
Setting $U=2V$ in this equation,  one obtains that $F(2V,V,\Delta_{\downarrow}, \Delta_{\uparrow})=F(2V,V,-\Delta_{\downarrow}, \Delta_{\uparrow})$, what is a condition for a first order transition, at which $\Delta_{\downarrow}$ changes discontinuously ($\Delta_{\uparrow}$ behaves continuously). 
Thus, for $U=2V$ and any temperature a transition occurs between solutions $(\Delta_{\downarrow}, \Delta_{\uparrow})$ and $(\Delta^{'}_{\downarrow}, \Delta_{\uparrow}) = (-\Delta_{\downarrow}, \Delta_{\uparrow})$ of  (\ref{eq:DeltaDown})--(\ref{eq:DeltaUp}), what coincides with the results obtained in Sec. \ref{sec:numresults} numerically. 
This transition is discontinuous if $\Delta_{\downarrow}\neq 0$ (a case of $T<T_{c}^{*}$ discussed in Sec. \ref{sec:phasediagHFA}) and it is continuous if $\Delta_{\downarrow}=0$ (a case of $T_{c}>T>T_{c}^{*}$ for $U=2V$).
Formally, such a continuous transition is not a transition and it corresponds rather to a smooth crossover between the phases, because the broken symmetries in both phases (here in the phases called as the AF and CO phases) are the same. 
Also, if $\Delta_{\downarrow}=0$ at $U=2V$, there is no specific features as expected for a second-order transition [cf. Fig. \ref{fig:HFAthermodynamicquantitiestwo}(c)].

Finally, let us analyze set (\ref{eq:DeltaDown})--(\ref{eq:DeltaUp}) for $U=2V$.
In such a case, the equation are decoupled (because $A=V\Delta_{\downarrow}$ and $B=2V\Delta_{\uparrow}$) and we can have two different temperatures $T_{c}$ and $T_{c}^{*}$ at which $\Delta_{\uparrow}$ and $\Delta_{\downarrow}$ vanish, respectively.
Assuming that $\Delta_{\uparrow}\rightarrow 0$ continuously at  $\beta_{c}=(k_{B}T_{c})^{-1}$, from (\ref{eq:DeltaUp}) one gets $\beta_{c}=2/V$.
This coincides with the order-disorder transition temperature determined in Sec. \ref{sec:HFA} at $U=2V$.
From (\ref{eq:DeltaDown}), one obtains equation for $\beta_{c}^{*}=(k_{B}T_{c}^{*})^{-1}$ as $I_{c}(\beta_{c}^{*}) = 1 / V$, where $I_{c}(\beta)$ is expressed by (\ref{eq:crittempIntegral}). 
That temperature corresponds to temperature at which the first-order transition line between two ordered phases ends, i.e., location of the isolated-critical point [cf. Figs. \ref{fig:HFAexphasediagram}, \ref{fig:HFAthermodynamicquantitiesone}(a), and \ref{fig:HFAthermodynamicquantitiestwo}(b)].

%\onecolumngrid
\bibliography{biblio_2_22}

\end{document}